# Dissipative particle dynamics simulation study on ATRP-brush modification of variably shaped surfaces and biopolymer adsorption

Samiksha Shrivastava[1], Ifra[2], Sampa Saha[2], and Awaneesh Singh[1*]

[1]Department of Physics, Indian Institute of Technology (BHU), Varanasi, India

[2]Department of Materials Science and Engineering, Indian Institute of Technology Delhi, New Delhi, India

**Abstract**

We present a dissipative particle dynamics (DPD) simulation study on the surface modification of initiator embedded microparticles (MPs) of different shapes via atom transfer radical polymerization (ATRP) brush growth. The surface-initiated ATRP-brush growth leads to the formation of a more globular MP shape. We perform the comparative analysis of ATRP-brush growth on three different forms of particle surfaces: cup surface, spherical surface, and flat surface (rectangular/disk-shaped). First, we establish the chemical kinetics of the brush growth: the monomer conversion and the reaction rates. We next argue the structure changes (shape-modification) of brush-modified surfaces by computing the radial distribution function, spatial density distribution, radius of gyration, hydrodynamic radius, and shape factor. The polymer brush-modified particles are well known as the carrier materials for enzyme immobilization. Finally, we study the biopolymer adsorption on ATRP-brush modified particles in a compatible solution. In particular, we explore the effect of ATRP-brush length, biopolymer chain length, and concentration on the adsorption process. Our results illustrate the enhanced biopolymer adsorption with increased brush length, initiator concentration, and biopolymer concentration. Most importantly, the flat surface loads more biopolymers than the other two surfaces when adsorption reaches saturation. The experimental results verified the same, considering the disk-shaped flat surface particle, cup, and spherical particles.

*Author for correspondence: awaneesh.phy@iitbhu.ac.in; awaneesh11@gmail.com

## 1. Introduction

The structural modification of MPs is gaining immense attention in the biomedical field (mainly in targeted drug/cell delivery, self-assembly, diagnostics)[1–4] due to their enormous applications. Recent studies show that MP's structure and shape significantly influence physiological interactions such as margination, circular half-life, and opsonization.[5–8] Apart from the extensively used spherical MPs, a noticeable rise has also been observed in MPs of various other shapes such as cup, disc, rod, tube, and ellipsoid as an emerging controlled drug delivery system.[9–12] Therefore, a controlled surface modification of the particles could be highly crucial to several biomedical applications.[13–17] There are numerous reports demonstrating the importance of surface modification of MPs such as polymeric cup-shaped particles using various techniques such as solvent evaporation,[18] vapor deposition,[19] mini-emulsion,[20,21] microfluidics.[22]

Out of numerous applications of surface-modified particles, the immobilization of biopolymers (e.g., enzymes) has been getting enormous attention in the last decade due to its use in biocatalysts, food processing, drug, and textile manufacturing, biosensors, and so on.[23–25] This inspired researchers to find a better solution to utilize biopolymers more efficiently to expand their applications.[16,26–29] Further, the immobilization of biopolymers to insoluble solid particles has enhanced the rapid arrest and controlled quenching of reaction kinetics. This is because of a comparatively more straightforward extraction of solid surface-bound biopolymers from the solution.[16,26–34] Thus, this immobilization process improves the stability and reusability of biopolymers against the change in pH, solvent, and temperature over successive chemical reactions.[35–37] The immobilization of biopolymers on solid surfaces has significant medical and industrial applications.[35] Additionally, surface-modified solid particles with polyelectrolyte ATRP-brush (e.g., poly(DMAEMA)) can immobilize a considerable amount of biopolymer through electrostatic adsorption in their brush morphology due to the high charge density.[37–39] Thus, brush-modified surfaces proved fascinating in immobilizing biopolymers due to their unique structures.[36,37]

Ifra and co-workers have recently illustrated differently-shaped MP modifications using the *electrojetting technique*.[14–16] In particular, different MPs are prepared from a blend of polylactide (PLA) and poly[methylmethacrylate-co-2-(2-bromopropionyloxy) ethyl methacrylate] (poly(MMA-co-BEMA)) in 3:1 ratio.[14] To fabricate the surface, poly(DMAEMA) brushes are synthesized via *grafting from* approach at the initiator (BEMA molecule) embedded surfaces via atom transfer radical polymerization (ATRP)[14–16] with DMAEMA monomer in the water ($pH \simeq 7$). The brush grafting density is regulated by altering the initiator concentration on the surfaces. The polymerization time yields ATRP brushes of different lengths and thus controls surface charge at the fabricated surface. Thus, the efficiency of electrostatic adsorption of biopolymer ($\alpha$-Glucosidase) onto the surface of a brush-modified particle gets enhanced.[16,36,37]

To understand the experimental findings by Ifra and co-workers[16], we present a DPD simulation study to gain further insights into the underlying physical phenomena at the atomic level. In DPD, the system evolution follows Newton's equation of motion.[40–43] Advantage of using the DPD technique is well-known for preserving the hydrodynamic behavior of modeled polymeric solutions, melts, and biopolymers in crowded and confined environments, which is crucial to estimating the system's dynamic behavior.[44,45] Hence, it is proved to be a very effective particle (DPD bead) simulation technique.[40,41] DPD is used extensively to model various types of free radical polymerization processes, and the effect of shape and size on the dynamics of the complex soft materials.[42,43]

This contribution presents a generic DPD model focusing on surface fabrication via surface-initiated ATRP-brush growth at differently-shaped MPs (such as the cup, sphere, and rectangular) surfaces and biopolymer adsorption occurring at these modified surfaces. Our simulation demonstrates a comparative study by varying the percentage of initiator concentration and polymerization time to vary the brush density and brush length and their effect on biopolymer adsorption. Further, we correlate them with the experimental findings.[15,16] The biopolymer is modeled as a linear free-rotating chain of soft beads connected by harmonic bonds.[42,43]

We organize this paper as follows. First, in section 2, we describe the computational model details and simulation parameters of the system. Then, section 3.1 discusses the results obtained for brush growth from the initiators implanted at the three different surfaces via surface-initiated ATRP in the solvent. Further, we discuss the biopolymer adsorption at the brush modified cup surface (CS), spherical surface (SS), and flat rectangular surface (RS) particles for various biopolymer concentrations in the solution. Next, section 3.2 presents the experimental results of CS, SS, disk surface modification, and biopolymer adsorption. Finally, in section 4, we conclude the paper with a summary.

## 2. Methodology and model parameters

**2.1. Dissipative particle dynamics.** DPD is a coarse-grained molecular dynamics (MD) approach where a bead symbolizes a molecule or a cluster of particles.[40,41] The beads interact through a soft-core potential, making DPD a more appropriate technique to simulate a system over a more considerable length and time scale than conventional MD simulations. Typically, DPD is utilized to simulate a system up to $100\ nm$ in the linear dimension and tens of microseconds for the time scale.[43,46–48] We briefly outline the key features of the DPD approach below. Moreover, a detailed description can be found in some original and latest publications.[40,47–50]

The equation of motion of each bead is governed by integrating Newton's second law:[40,41]

$$\frac{d\vec{r}_i}{dt} = \vec{v}_i; \qquad m_i \frac{d^2\vec{r}_i}{dt^2} = \vec{f}_i(t), \qquad (1)$$

where $\vec{r}_i, \vec{v}_i, m_i$, and $\vec{p}_i = m_i \vec{v}_i$ denote the position, velocity, mass, and momentum related to $i^{th}$ bead. The total force $\vec{f}_i = \sum_{j \neq i} \vec{F}_{ij} = \sum_{j \neq i}(\vec{F}_{ij}^C + \vec{F}_{ij}^D + \vec{F}_{ij}^R)$ acting on $i^{th}$ bead due to all other $j$ beads within an interaction distance $r_c$ consists of three pairwise additive forces, where $\vec{F}_{ij}^C, \vec{F}_{ij}^D$, and $\vec{F}_{ij}^R$ represent the conservative, dissipative and random force. The force symmetry $\vec{F}_{ij} = -\vec{F}_{ji}$ ensures the momentum conservation. The cutoff distance $r_c$ is considered as an intrinsic length scale in the DPD model.[40,41,49] Each bead in the system has equal mass, i.e., $m_i = m$. The typical conservative force considered in DPD is[40]

$$\vec{F}_{ij}^C = a_{ij} \omega_C(r_{ij}) \hat{r}_{ij}, \qquad (2)$$

where $\omega_C(r_{ij}) = (1 - r_{ij}/r_c)$ is the weight function that belongs to a soft repulsive interaction which enables a more considerable length and time scale in DPD; $a_{ij}$ denotes the maximum repulsion parameter between $i$ and $j$ beads; $r_{ij} = |\vec{r}_{ij}|, \hat{r}_{ij} = \vec{r}_{ij}/|\vec{r}_{ij}|$, and $\vec{r}_{ij} = \vec{r}_i - \vec{r}_j$.

The dissipative force, which represents the effects of viscosity, and the random force, which denotes the impact of thermal fluctuations, are described by[40,41]

$$\vec{F}_{ij}^D = -\gamma \omega_D(r_{ij})(\hat{r}_{ij} \cdot \vec{v}_{ij}) \hat{r}_{ij}, \qquad (3)$$

and

$$\vec{F}_{ij}^R = \sigma \omega_R(r_{ij}) \sqrt{\Delta t} \xi_{ij} \hat{r}_{ij}. \qquad (4)$$

Here, $\gamma$ and $\sigma$ denote the strength of dissipative and random forces, $\vec{v}_{ij} = \vec{v}_i - \vec{v}_j$. $\Delta t$ is the simulation time step. The noise amplitude $\xi_{ij}$ in Eq. (4) signifies a Gaussian random distribution with zero-mean: $\langle \xi_{ij}(t) \rangle = 0$, and unit variance:[40,41] $\langle \xi_{ij}(t) \xi_{kl}(t') \rangle = (\delta_{ik} \delta_{jl} + \delta_{il} \delta_{jk}) \delta(t - t')$. The symmetry relation $\xi_{ij} = \xi_{ji}$ and the forces, $\vec{F}_{ij}^D$ and $\vec{F}_{ij}^R$ act along the line of bead centers, ensuring momentum conservation. For a correct canonical equilibrium state, the dissipative and random forces are coupled with the following relations:[40,41] $\sigma^2 = 2\gamma k_B T$, and $\omega_D(r_{ij}) = \omega_R(r_{ij})^2 = (1 - r_{ij}/r_c)^s$ for $r_{ij} < r_c$ where $k_B$ is the Boltzmann constant, and $T$ is the equilibrium temperature of the system. Here, we took the traditional weight function exponent $s = 2$. However, other choices are also allowed if the above two conditions are fulfilled.[49,51] Overall, the notable advantage of the DPD approach is that the pairwise forces shown in Eqs. (2-4), conserves the momentum locally, thus upholding the correct hydrodynamic[52] behavior with only a few hundred particles.[40,41]

We set the interaction parameter between the compatible beads, $a_{ij} = 25$ (in reduced DPD units). This choice is based on the compressibility of water by coarse-graining ten water molecules into one bead.[40] In general, for any two incompatible beads, we set $a_{ij} = 60$.[43] We provide the other $a_{ij}$ values in a short while. The parameters $r_c$, $m$, and $T$ are set to 1.0 and $\gamma = 4.5$ in reduced DPD units. The typical energy scale of the model is set by $k_B T$.[40,53] We use the modified velocity-verlet algorithm to integrate the equations of motion.[40,54] The simulation time step is set to $\Delta t = 0.02\tau$ where $\tau = (mr_c^2/k_B T)^{1/2}$ is defined as the characteristic time scale. The total bead number density in the simulation box is set to $\rho = 3$, which is a reasonable choice for DPD simulation of liquids.[40,41] The above choices of coarse-graining yield the dimensional values of length $r_c \approx 0.97\ nm$, and time $\tau \approx 8.3\ ps$.[55,56]

We simulate the polymer chains as a bead-spring model[56] where the harmonic bond potential $E_b = 1/2\ k_b(r - r_0)^2$ connects neighboring DPD beads of the chain. Here $k_b = 128$ is the elastic bond strength, and $r_0 = 0.5$ is the equilibrium bond distance.[57] The angle potential $E_a = 1/2\ k_a(\cos\theta - \cos\theta_0)^2$ provides the polymer chain stiffness. We set the potential strength $k_a = 5$, and the equilibrium angle $\theta_0 = 180$ degrees;[56–58] $\theta$ denotes the angle between two successive bonds along a polymer chain.

**2.2 Modeling the nanoparticles.** In the framework of DPD simulation, we model the differently shaped MPs (CS, SS, and RS) as rigid bodies. Therefore, the forces and torques acting on rigid MPs are computed as the sum of all the forces and torques on constituent DPD beads. The particles are made of PLA blended with poly(MMA-*co*-BEMA) in a 3:1 ratio, as described in the experiments,[16]. The poly(BEMA) compound acts as an initiator to instigate the ATRP process with DMAEMA monomers available in the solution to fabricate the surface. The cup (half-sphere) and spherical-shaped MPs are composed of double-layered DPD beads arranged on the vertices of a geodesic grid generated by subdividing an icosahedron.[59] The inner and outer radii are set to $R_{in} = 7.7r_c$, $R_o = 8.0r_c$. The cup-shaped MP contains $N_{cs} = 1962$ beads ($\phi_{cs} = 5.23 \times 10^{-3}$) and the spherical-shaped MP comprise $N_{ss} = 3924$ beads ($\phi_{ss} = 1.04 \times 10^{-2}$). The rectangle-shaped MP surface is also modeled as a double-layered flat structure with $N_{Rs} = 1960$ beads ($\phi_{rs} = 5.23 \times 10^{-3}$) arranged in a regular lattice structure with a lattice constant of $0.5r_c$. Each layer contains an equal number of beads. The rectangular MP size is $17.5r_c \times 14r_c$ (in reduced DPD units) with $dz = 0.5r_c$ dimensionless unit in thickness. The number density of beads is kept sufficiently high ($\rho_m \sim 11$) in each layer of CS, SS, and RS to prevent the penetration of ATRP-brush and other beads.

The experimental studies[14–16] illustrated that only a fraction of poly(BEMA) molecules on the MP's surface act as initiators in the solution. Our simulation models this experimental observation by considering a certain fraction of MP beads as the initiator beads. Furthermore, we chose the initiator beads randomly at either particle surface for the CS and RS, whereas for SS, they are selected only on the outer surface. Note that the flat surface is a disk-shaped particle in the experiment, whereas we model it as a rectangular-shaped particle in the simulation. We considered three initiator values: $N_i = $ 20, 50, and 100, *i.e.*, 1%, 2.5%, and 5% of MP beads, respectively; corresponding volume fractions are $\phi_i = 5.3 \times 10^{-5}$, $1.33 \times 10^{-4}$, and $2.66 \times 10^{-4}$. Thus, polymer brushes ($B$) will grow from the surface during polymerization. The MPs are insoluble in the solvent ($s$),[16] we set their interaction with the solvent $a_{ms} = 45$. To model the fact that the interaction between MP bead ($m$) and monomer bead ($M$) is slightly incompatible, we set $a_{mM} = 35$, and its interaction with the polymer brush is set as $a_{mB} = 35$.

**2.3 Surface fabrication of microparticles via ATRP-brush.** To simulate ATRP-brush growth, we exploit the recently developed DPD approach for *living radical polymerization*, where a set of relevant elemental reactions model the ATRP processes.[42,43,60] The elements involved in ATRP are initiator and monomer; all are modeled as DPD beads. The main reaction steps are the initiation and propagation of monomers in the explicitly considered solvent beads.

The polymerization in our approach is similar to the earlier modeled via coarse-grained MD and MC simulations.[61,62] Herein, we picked the monomer volume fraction, $\phi_M = 1.0 \times 10^{-1}$ ($N_M = 37500$) to grow up the polymer brush. We randomly selected initiator beads to begin the ATRP reactions. In each reaction step, we randomly select monomer beads within the interaction radius, $r_i = 0.7 r_c$ of each initiator bead.[42,43] The selected bead pair will form a covalent bond with the polymerization probability $0 < P_r^x < 1$, where superscript $x$ stands for the type of chemical reaction; $x = i$ for the initiation, and $x = pM$ for the propagation with monomers.[63,64] Next, we draw a random number $n_r \in (0,1)$ from a uniform distribution and compare it with the reaction probability $P_r^x$. The covalent bond formation reaction step is accepted when $n_r < P_r^x$, otherwise rejected. Thus, each successful reaction step yields an irreversible covalent bond represented by the harmonic bond potential with $k_b = 128$, $k_a = 5$, and $\theta_0 = 180$ (in reduced DPD unit).

In our simulation, we set the initiation probability $P_r^i = 0.5$, and the chain propagation probability $P_r^{pM} = 0.05$.[42,63,64] The respective reaction rate constants can be adjusted efficiently by selecting different values of $P_r^x$. Nevertheless, lower $P_r^x$ values are chosen to ensure *kinetically controlled* polymerization. The stated value of $r_i$ is chosen to reproduce the expected linear first-order kinetics for living radical polymerization. Any smaller values delay the brush growth, and higher values significantly deviate from the linear first-order kinetics.[43] We exclude the degenerative chain transfer and the termination reaction of active radicals to dormant species in ATRP due to the characteristics of living polymerization.[65,66] The details of bead types involved in the primary reaction steps are summarized via a schematic shown in Fig. 1. Since the characteristic simulation time step is set to $\Delta t = 0.02\tau$, and the reaction time interval between any two successive reaction steps is selected to be $\tau_r = 0.2\tau$,[42,67] the reactions are performed every ten simulation time steps.[42]

The monomer and ATRP-brush beads are considered hydrophilic (compatible) with solvent beads, and hence, their interaction parameters are set to $a_{MS} = a_{BS} = 25$ (in DPD units). Similarly, for monomer and brush (polymerized monomer) bead interaction, $a_{MB} = 25$. The experimental results[14,15] suggest that in a pure solvent (water at $pH \approx 7$), poly(DMAEMA) brushes are positively charged due to protonated amino groups present in the brush. However, the explicit use of charge distribution and corresponding electrostatic forces in the DPD simulation approach of biological systems is rare.[68–70] Due to the charge screening, electrostatic interaction becomes short-range. Therefore, to incorporate the effect of repulsive interaction between the brush beads carrying positive charges, we choose $a_{BB} = 27$.

**2.3 Biopolymer adsorption at the fabricated nanoparticle surface.** We modeled the biopolymer ($\alpha$-glucosidase enzyme) in the same way as the brush chain- a linear chain of $N_b$ DPD beads, connected by harmonic bonds with an elastic constant $k_b = 128$, and equilibrium bond distance $r_0 = 0.5$.[71] The angle potential coefficient between two consecutive bonds along with the biopolymer chain $k_a = 5$, and $\theta_0 = 180$ is an equilibrium angle between the successive bonds. The mass of each biopolymer bead is set to 1.0 in the reduced DPD unit. We study biopolymer adsorption for three different concentrations: $c_b = 2\%N$ ($\phi_b = 2.0 \times 10^{-2}$), $4\%N$ ($\phi_b = 4.0 \times 10^{-2}$), and $8\%N$ ($\phi_b = 8.0 \times 10^{-2}$). The corresponding number of DPD beads are $N_b = 7500$, 15000, and 30000, respectively. To study the effect of biopolymer length ($l_b$) on the adsorption at fabricated surfaces, we choose $l_b = 25$, 10, and 1. The biopolymer ($\alpha$-glucosidase[16]) is chemically compatible with the solvent and negatively charged molecule. Therefore, the positively charged brush beads will attract the negatively charged biopolymers. Thus, to incorporate this favorable interaction, we set the interaction parameters between these beads, $a_{Bb} = 15$, and between the biopolymer beads, $a_{bb} = 27$.

We consider a simulation box of size $50r_c \times 50r_c \times 50r_c$. The total number density in the box is fixed at $\rho = 3r_c^{-3}$ (i.e., $N = 375000$ beads).[40–43] The periodic boundary conditions are applied in the $x$- and $y$-directions. In contrast, the simulation box in the transverse $z$-direction is bounded by amorphous solid walls.[42] The height of both walls is fixed to $h = 1$, and bead density is set $\rho_w = 3$ (volume fraction, $\phi_w = 4.0 \times 10^{-2}$). In addition, we apply bounce-back boundary conditions at the fluid-wall interface[72] to inhibit the penetration of solvent, moieties, and polymer brush beads into the walls.[73] We consider repulsive interaction ($a_{wj} = 60$) between wall beads and other beads in the system. The solvent volume fraction in our simulation box is held fixed to $\phi_s \approx 8.3 \times 10^{-1}$ (for $\phi_b = 2.0 \times 10^{-2}$), $8.1 \times 10^{-1}$ (for $\phi_b = 4.0 \times 10^{-2}$), and $7.7 \times 10^{-1}$ (for $\phi_b = 8.0 \times 10^{-2}$). The interaction parameters for all DPD beads are summarized in Table 1.

To start the DPD simulation, we place the microparticle in a simulation box and generate the initial configuration of monomer and solvent by randomly placing them in the box. Then, we equilibrate the system for $t = 1 \times 10^5$ simulation time steps, and allow the polymerization reactions to begin. First, we let the ATRP brush grow up to three different time steps: $t_{BG} = 400, 800,$ and $1500$ to get the fabricated MP surface with varying brush lengths. Next, we introduced the biopolymers in the solution containing fabricated MPs and observed the biopolymer adsorption up to $t_{bA} = 3000$ simulation time steps.

## 3. Results and discussion

### 3.1. Simulation Study

We utilize the DPD simulation approach to study the biopolymer adsorption at ATRP-brush modified different-shaped MPs such as cup, spherical, and rectangular surfaces. To begin with, first, we study the surface modification of a cup particle where initiators are equally distributed at the inner and outer surfaces. Three initiator percentages $c_i = 1\%, 2.5\%, 5\%$ of $N_{cs}$ are considered to differentiate its effect on surface fabrication. We allow the atom transfer radical polymerization (ATRP) reaction to begin with monomers in the solution. To analyze the polymerization kinetics, we plot the monomer conversion ($\text{Conv}_M = [M]_{rt}/[M]_0$) and the reaction rate, $\log([M]_0/[M]_{ut})$ as a function of time ($t_{BG}$) in Fig. 2, where $[M]_0$ denotes the initial monomer concentration, $[M]_{rt}$ and $[M]_{ut}$ indicate the reacted and unreacted monomer concentration, respectively, at $t_{BG}$. The polymerization is allowed up to $t_{BG} = 1500$ for all three initiator concentrations. As illustrated in Fig. 2(a), with an increase in $c_i$, $\text{Conv}_M$ increases linearly, which yielded nearly 6% conversion for $c_i = 1\%$ (black curve), 13% conversion for $c_i = 2.5\%$ (red curve), and 21% conversion for $c_i = 5\%$ (green curve) within the given simulation time. Figure 2(a) demonstrates that, in ATRP, polymer chains grow linearly with the monomer conversion. Thus, a higher monomer conversion corresponds to a longer brush length.

The reaction rate, $\log([M]_0/[M]_{ut})$ is also varying linearly for all $c_i$ values illustrated in Fig. 2(b). Thus, confirming the first-order reaction kinetics of ATRP, as expected for a living/controlled polymerization with constant free radical concentrations during the reaction process.[42,63,64] Figures 2(c-e) show the snapshots of fabricated CS for the ATRP-brush growth till $t_{BG} = 1500$ at different initiator concentrations. Figures 2(f-h) display brush density variation around CS in the $xy$-plane at $l_z = 25$ corresponding to Figs. 2(c-e). The yellow circle shows the crosssection of CS, and the red region displays the ATRP-brush density. These snapshots clarify that the brush density is higher for higher monomer conversion (~21%) for $c_i = 5.0\%$ at CS. The presence of the red region within CS (yellow circle) also confirms the initiators' presence at the inner surface of CS.

The two most frequently used physical parameters that characterize the size of a macromolecule are the radius of gyration ($R_g$) and hydrodynamic radius ($R_h$). Both $R_g$ and $R_h$ used different ways to compute the size of a macromolecule and arrive at a similar conclusion. We calculate the radius of gyration[74,75] $R_g = \left(\frac{1}{N}\sum_i \langle r_i^2 \rangle\right)^{1/2}$ as a function of time $t_{BG}$ for $c_i = 1.0\%, 2.5\%,$ and $5.0\%$, as displayed by the black, red, and green curves in Fig. 3(a). Here $N$ denotes the total number of CS and brush beads, and $r_i$ represents the distance of $i^{th}$ bead from the center of mass of CS and brush beads. The angular brackets signify the averaging over five ensembles. The plots in Fig. 3(a) show that brush fabricated cup particle size increases with time as the degree of polymerization ($\text{Conv}_M$) increases for all $c_i$. However, $\text{Conv}_M$ is more for higher $c_i$ (= 5.0%) at a set duration of polymerization. Hence, the larger values of $R_g$ is observed as a function of time (see the curves in Fig. 3(a)).

Note that in ATRP, one monomer adds up at a time to a growing chain (i.e., the propagation rate is the same for each initiator). Consequently, the growing chains are nearly monodispersed.[62,64–66] Therefore, at a given monomer concentration, the individual chain length (degree of polymerization) is longer for a low initiator concentration for any given time during the polymerization. Thus, from the results shown in Fig. 2, the estimated individual brush length is $l_B \simeq n_B r_0$ where $n_B \simeq 115, 100,$ and $86$ are roughly the number of polymerized beads per brush chain for $c_i = 1.0\%, 2.5\%,$ and $5.0\%$, respectively, at $t_{BG} = 1500$. Given the fact that polymer brushes are repulsive with CS ($a_{mB} = 35$) and also with themselves ($a_{BB} = 27$) due to positively charged DPD beads, the ATRP-brushes are more swollen and thus relatively more extended in the solvent for $c_i = 5.0\%$ than for $c_i = 1.0\%$ and hence, higher $R_g$ is observed.

We compute $R_h$ to gain further insight into the brush-modified particle size. $R_h$ imitates the size of a solvated molecule more closely. Therefore, it is a more appropriate biological parameter to determine the size of a molecule in the context of its environment. Thus, $R_h$ of a modified particle is measured by assuming that the brush embedded on its surface moves through the solution and is resisted by the solvent viscosity. The hydrodynamic radius formulation is $\frac{1}{R_h} = \frac{1}{2N^2}\left\langle \sum_{i \neq j} \frac{1}{r_{ij}} \right\rangle$,[74–77] where $r_{ij}$ is the distance between $i$ and $j$ beads of polymer brush and CS ($N$); the angular brackets denote the averaging over five ensembles. Like $R_g$ in Fig. 3(a), $R_h$ also increases linearly with time as the brush density increases at CS, having almost the same order of magnitude. The brush swells more with higher initiator concentration, thus, explaining a higher $R_h$ for $c_i = 5.0\%$.

The ratio of $R_g$ and $R_h$, $\rho_{sf} = R_g/R_h$ (also known as the shape factor), characterizes the shape of a macromolecule. The shape factor for a globular polymer structure is approximately $\rho_{sf} \sim 0.775$.[74,75] However, any departure from the globular to nonglobular (or elongated) shape, $R_g/R_h$ attains higher values as $R_g$ turn out to be larger than $R_h$. For a hollow sphere, $\rho_{sf} \sim 1.0$. At early times, the brush growth is minimal, and hence, the shape factor for the modified CS at this stage is estimated as $\rho_{sf} \in (1.05 - 1.065)$ for all $c_i$ values as displayed in Fig. 3(c). However, ATRP-brush modified CS tends toward attaining a relatively more globular shape with time; thus, $\rho_{sf}$ decreases. At late times, the shape factor is smaller ($\rho_{sf} \simeq 0.985$) for $c_i = 5.0\%$ than for the other two $c_i$ values (see Fig. 3(c)). Nevertheless, when we allow the brush growth for a much longer time at which $\text{Conv}_M \rightarrow 90\%$, the fabricated MP attains the shape factor $\rho_{sf} \simeq 0.94$ (this result is not displayed here due to brevity).

To illustrate the polymerized (brush) beads distribution tethered, we compute the radial distribution function (RDF), $g(r) = \rho_{lB}/\rho_B$ of brush beads at a radial distance, $r = (|dx|^2 + |dy|^2 + |dz|^2)^{1/2}$ from the CS beads. Here, $\rho_{lB}(r) = n_{lB}(r)/V_{sh}$ denotes the local brush density, and $\rho_B = N_B/V$ indicates the total brush density. $n_{lB}(r)$ is the local number of brush beads (displayed in the blue and green colors) in a shell of volume $V_{sh}$ at a distance $r$ from the microparticle bead (displayed in the orange color), $N_B$ represents the total brush beads, and $V$ is the total volume of the box. The plots in Fig. 4(a) exhibit the variation in RDF for three different $c_i$ values as denoted by the black, red, and green symbols. As expected, the larger peak width and height shown by the green curve validate that brush beads are more tightly bound around CS at $c_i = 5.0\%$ due to higher monomer conversion than at $c_i = 1.0\%$ and 2.5%. The plot of number density $\rho(z)$ of brush beads along the transverse ($z$)-direction is displayed in Fig. 4(b). The curves at different $c_i$, further certify that more brush beads are localized around CS ($10 < z < 30$) at $c_i = 5.0\%$ than at $c_i = 1.0\%$ (black curve) and $c_i = 2.5\%$ (red curves). Overall, an increase in the initiator concentration yields noticeable brush swelling, which could facilitate the biopolymer diffusion into the brush matrix, thereby enabling immobilization.

We next consider SS and RS to study the surface modification due to ATRP-brush growth and its influence on biopolymer adsorption. However, first, we compare corresponding physical properties related to the structural change of SS and RS with CS, as discussed above. Then, since the brush modified CS is highly swollen for the brush growth up to $t_{BG} = 1500$ at $c_i = 5.0\%$, we compare the SS and RS fabrication with CS by considering the polymerization up to the same simulation time steps for $c_i = 5.0\%$.

In Fig. 5, we plot $\text{Conv}_M$ and $\log([M]_0/[M]_{ut})$ as a function of time for SS and RS and compared them with the results obtained for CS. The initiators are evenly distributed at both the surfaces of RS and only at the outer surface of SS. The surface fabrication is led by growing the ATRP brushes up to $t_{BG} = 1500$. We observe that $\text{Conv}_M$ for SS (red curve) is slightly higher than RS (green curve) and CS (black curve) at early times ($t_{BG} \lesssim 350$). Whereas the gap is enhanced over time, as displayed in Figs. 5(a)-(b), due to higher monomer accessibility to active radicals tethered at a larger outer surface of SS than CS and RS. The propagation reaction ($\text{Conv}_M$) for RS is also growing at almost the same rate as fS at early times ($t_{BG} \lesssim 1000$). However, a slightly higher conversion is obtained for RS beyond this time. For RS, the monomers are equally accessible to initiators at both surfaces. Thus, initiators/active radicals have equal accessibility to monomers. The brush growth at the outer CS can occur at the same rate for the SS; however, the monomers are less accessible to initiators at the inner concave surface of CS. Thus, the effective rate of chain propagation at CS is balanced such that it yields less monomer conversion than on RS and SS. Overall, Fig. 5(b) illustrates that the propagation reaction rate ($\log([M]_0/[M]_{ut})$) at different surfaces varies linearly with time ($t_{BG}$) thus, exhibiting the expected first-order reaction kinetics.

We illustrate the morphology of fabricated CS, SS, and RS due to ATRP-brush growth up to different polymerization times, $t_{BG} = 400$ (Fig. 6(a-c)), 800 (Fig. 6(d-f)), and 1500 (Fig. 6(g-i)) at $c_i = 5.0\%$. The $\text{Conv}_M$ at CS, SS, and RS for $t_{BG} = 1500$ are $\sim 23\%, \sim 26\%,$ and $\sim 24\%$, respectively, as illustrated in Fig. 5(a). The corresponding estimated brush length is $l_B \simeq n_B r_0$ where $n_B \simeq 79, 97$, and $86$ are the number of polymerized beads per brush chain bounded at CS, SS, and RS, respectively. At $t_{BG} = 400$, the monomer conversion is almost the same at all surfaces, hence, nearly the same brush length: $l_B \simeq 21 r_0$ for CS and RS, and $l_B \simeq 26 r_0$ for SS are assessed. However, $l_B$ increases with time and is slightly different from each other at different surfaces due to diverse monomer conversion, as can be seen in Fig 5. For the polymerization up to $t_{BG} = 800$, the approximate brush lengths are $l_B \simeq 41 r_0, 53 r_0$, and $43 r_0$ at CS, SS, and RS where $\text{Conv}_M \simeq 11\%, 14\%$, and $12\%$, respectively. Figures 6(j-l) show the brush density variation around MP surfaces in the $xy$-plane at $l_z = 25$, corresponding to Figs. 6(g-i). The yellow color represents the crosssection of various MPs, and the red region displays the ATRP-brush density.

The shape and size of ATRP-brush modified differently-shaped particles are compared at $c_i = 5.0\%$. First, we plot $R_g$ as a function of $t_{BG}$ in Fig. 7(a). The difference in $R_g$ values are due to different MP structures; $R_g$ for SS is higher than CS and RS. However, the latter two surfaces maintained similar growth in $R_g$ against time with a tiny advancement for RS at late times due to higher $\text{Conv}_M$ (shown in Fig. 5). With a considerable $\text{Conv}_M$ at SS (as discussed in Fig. 5), the initial surge in $R_g$ is maintained till late times. Similar to $R_g$, the hydrodynamic radius ($R_h$) for the modified surfaces varies linearly with time, as presented in Fig. 7(b). In Fig. 7(c), we plot the shape factor ($\rho_{sf}$) to characterize and compare the temporal variation in the shape of modified MPs. At early times, when $\text{Conv}_M$ is small, $\rho_{sf}$ is highest for RS and lowest for SS due to their different unmodified shapes. However, $\rho_{sf}$ for CS and RS approach a similar value as the polymerization increases with time. Though, $\rho_{sf}$ for SS remains the lowest for the entire polymerization period considered here.

We compare $g(r)$ versus $r$ for the ATRP brushes around CS in Fig. 8(a), SS in Fig. 8(b), and RS in Fig. 8(c) at $c_i = 5.0\%$ for three different brush lengths assessed at $t_{BG} = 400, 800,$ and $1500$. The corresponding brush lengths at each surface are estimated and discussed a short while ago. Figure 8 demonstrates that the peak height increases with brush length, and its position moves to higher $r$ for the modified particles. This suggests an increase in the local brush density ($\rho_{lB}$). Thus, the brushes are more closely bound to MP surfaces. The increase in the width of RDF with $r$, validates the spreading of growing brush chains. The local brush density becomes negligible with $r \to \infty$ and hence, $g(r) \to 0$. Following the previous observations as depicted in Figs. 5-7, the variation of $g(r)$ is very similar for both CS and RS; nevertheless, $g(r)$ for RS has a slight edge (in Fig. 8(c)) over CS (in Fig. 8(a)). Figure

8(b) shows that the peak of $g(r)$ for SS is lower than for CS and RS. The reason for this could be the lower $n_{lB}$ per shell because of the hollow core of the spherical MP and higher $N_{SS}$ (= $2 \times N_{CS}$ or $2 \times N_{RS}$), which reduces $\rho_{lB}(r)$. Since SS has a hollow core, and no other beads, including the brush beads, can penetrate it, we also see a sudden jump at $r \approx 2R_0$.

Recall that the brush grafting density increases with the polymerization, which causes an increase in the surface charges. Thus, the brushes get stretched out and swollen due to the incompatible interaction between the brush beads ($a_{BB} = 27$). Furthermore, the negatively charged biopolymers ($a_{bb} = 27$) pave the way to enter into the brush region and get adsorbed due to much favorable interaction ($a_{Bb} = 15$) between the swollen brush and biopolymers.

We introduce the equilibrated biopolymer chains of length $l_b = 25$ in the solution, and analyze the adsorption at different brush modified surfaces for a period $t_{bA} = 3000$. In Figs. 9(a-c), we compare the fraction of biopolymer adsorbed ($\phi_{bA}$) with time on fabricated CS (black curve), SS (red curve), and RS (green curve) in the presence of various fractions of biopolymers, $\phi_b = 2 \times 10^{-2}$, $4 \times 10^{-2}$, and $8 \times 10^{-2}$, respectively. We found that the biopolymers are getting adsorbed within the brush region at nearly the same rate on different brush-modified surfaces for some initial period ($t_{bA} \lesssim 300$). However, as time progresses, $\phi_{bA}$ for SS is growing faster than CS and RS for up to $t_{bA} \lesssim 900$. This observation is consistent with all the biopolymer concentrations used in the simulation. Furthermore, it seems pretty evident as the brush length ($l_B$) and the radius of gyration ($R_g$) are prominent for SS due to higher monomer conversion than CS and RS.

To compute $\phi_{bA}$, we count the number of biopolymer beads within $R_g$ of modified MPs and then scale it with the total number of beads ($N$) in the system. Interestingly, for $t_{bA} > 900$, we find a gradual crossover in $\phi_{bA}$ curves for RS to its corresponding noticeable higher value (see the green curves in Fig. 9) than for the other two surfaces at all the biopolymer concentrations ($\phi_b$). The crossover becomes more prominent with increasing values of $\phi_b$ as depicted in Figs. 9(b-c). On the other hand, $\phi_{bA}$ data for SS (red curve) and CS (black curve) converge to nearly the same lower value (steady-state value) for a given $\phi_b$ at late times. However, the red curve ($\phi_{bA}$ for SS) approaches the steady-state faster than the black curve (for CS) at a relatively higher $\phi_b$. At late times, the reason for a lower $\phi_{bA}$ at SS could be a relatively higher volume of the swollen brush region due to a more considerable brush length ($l_B$) caused by higher monomer conversion at SS. The larger volume implies a more porous brush region around SS. The biopolymers more easily diffuse into the brush matrix due to a favorable interaction ($a_{Bb} = 15$) between the brush and biopolymer beads. Note that biopolymers are negatively-charged molecules. Therefore, some biopolymers can easily diffuse out of the brush region after relatively quicker initial adsorption due to unfavorable interaction ($a_{bb} = 27$) between the biopolymer beads. Therefore, the number of trapped biopolymers within fabricated SS is relatively smaller than RS.

To demonstrate the effect of brush lengths on biopolymer adsorption, we consider the surface-initiated ATRP-brushes grew up to $t_{BG} = 400$, 800, and 1500 at each surfaces considered here for $c_i = 5\%$ and $\phi_b = 8 \times 10^{-2}$. The corresponding brush lengths are $l_B \approx 21, 41$, and 79 for CS, $l_B \approx 26, 53$, and 97 for SS, and $l_B \approx 21, 43$, and 86 for RS (brush lengths are measured in the units of $r_0$). To analyze the biopolymer adsorption at different brush lengths, we first plot $g(r)$ versus $r$ in Fig. 10 at $t_{bA} = 3000$, mainly to show the radial distribution of biopolymers within the brush region. Here, we take the biopolymer chain length $l_b = 25$. In ascending order, the black, red, and green curves exhibit $g(r)$ data for different brush lengths. Figure 10(a) displays that for CS, $g(r)$ peak strength and position enhanced, and also it is getting wider with increasing $l_B$. This implies that a more fraction of biopolymers have uniformly diffused into the brush region for a longer brush length; a similar behavior is observed for SS and RS, as shown in Figs. 10 (b-c). A sudden jump in $g(r)$ is observed for SS at $r \approx 2R_0$ due to the absence of biopolymer molecules within the hollow SS, as explained earlier in Fig. 8(b). We plot $\phi_{bA}$ versus $t_{bA}$ in Figs. 10(d-f), which displays that biopolymer adsorption increases with $l_B$ as shown by the black, red, and green curves. Similar to the results in Fig 9, the fraction of biopolymer adsorbed is relatively more for RS as in Fig. 10(f) than CS and SS as in Figs. 10(d-e). After an initial growth in $\phi_{bA}$, data saturates to a specific value for each brush length at late times; this behavior remains the same for all the modified surfaces. The adsorption is relatively minor for the shorter brushes, and the corresponding data saturates early.

The biopolymer length ($l_b$) significantly influences the amount of adsorption on brush-modified surfaces, as illustrated in Fig. 11. To characterize it, we consider biopolymers of three different chain lengths $l_b = 1, 10$, and 25, and monitor their adsorption up to $t_{bA} = 3000$ at $\phi_b = 8 \times 10^{-2}$. The plots of $g(r)$ versus $r$, and $\phi_{bA}$ versus $t_{bA}$ curves are displayed in Fig. 11 with the black, red, and green symbols. They exhibit the biopolymers distribution, and the fraction of it adsorbed within the ATRP-brush matrix on CS, SS, and RS. The brushes are allowed to grow on these surfaces till $t_{BG} = 1500$ for $c_i = 5\%$. These plots explain that the adsorption is more for the longer biopolymer chains, independent of the modified surface type. Notice the RDF curves in Figs. 11(a-c), the higher peak strength at lower $r$ certifies that biopolymers are more closely packed around the modified surfaces for $l_b = 25$ and $10$ (see the green and red curves). However, the black $g(r)$ curves show that when we coarse-grain the biopolymer with a single bead ($l_b = 1$), they are loosely bounded around the modified surfaces and uniformly distributed in a wide range of $r$ from the surface beads.

The curves in Figs. 11(d-f) demonstrate that $\phi_{bA}$ saturates to a finite value, i.e., the biopolymer adsorption attains a steady-state value, for all the cases studied here within the adsorption time $t_{bA} = 3000$. Similar to the results obtained for $l_b = 25$ (see Figs. 9 and 10), $\phi_{bA}$ is higher at RS ($\phi_{bA} \approx 11.6 \times 10^{-3}$) than CS and SS with $\phi_{bA} \approx 10.6 \times 10^{-3}$ for $l_b = 10$. It is interesting to note that $\phi_{bA}$ reaches its steady-state (constant) value sooner with decreasing biomolecule size $l_b$. The gap in $\phi_{bA}$ values at different modified surfaces are also getting insignificant. Therefore, for $l_b = 1$, $\phi_{bA} \approx 6.0 \times 10^{-3}$ for all the modified surfaces (see the black curves in Fig. 11 (d-f)). This could be attributed to a quick diffusion of smaller molecules into and out of the brush region as the diffusion coefficient $D \sim N_b^{-1}$ where $N_b$ is the number of beads in a biopolymer chain. Since the diffusion of longer biopolymer is slow, its saturation takes more time than smaller chains. However, when longer biopolymer chains are diffused into the brush matrix due to strong favorable interaction, they mostly remain trapped due to structural constraints.

### 3.2. Experimental Study

Further, we performed an additional experimental study to verify the simulation results. We focus on biopolymer adsorption at differently shaped brush modified surfaces such as cups, spheres, and disc surfaces; Recall that we have considered the rectangular surface in our simulation instead of the disc-shaped surface. Nevertheless, we presumed that both the flat surfaces (rectangular or disc) would regard the same result within the statistical error. The detail of the experimental setup is briefly provided in SI. However, we can find a more detailed design and explanation in some recent publications.[14,16]

In the experimental study, we synthesize poly(DMAEMA) brush-modified particles of various shapes such as spheres, cups, and discs following our previous publications[14,16] to study the effect of particles' shape on enzyme immobilization efficiency. To serve the purpose, first, we copolymerize the acrylate monomers, i.e., methyl methacrylate (MMA) and 2-hydroxy ethyl methacrylate (HEMA), and modify the hydroxyl groups of HEMA moieties via reacting with bromopriopionyl bromide (poly(MMA-co-BEMA)) as discussed in our previous study.[14,16] The bromo functional moiety of the copolymer is then utilized as an ATRP initiator to grow polymer brushes from the surface of the particles consisting of the copolymer,

poly(MMA-co-BEMA). The particles with different shapes are made by electrojetting a blend of polylactide (75%) and poly(MMA-co-BEMA)(25%) under various conditions, as stated in Table S1.[14,16]

During the electrojetting process, the polymer droplets are thought to be attracted to the collector at a low solution concentration. The polymer droplet starts bending depending upon the process parameters, resulting in the cup-shaped particle formation. However, in the case of high concentrations ($3w/v\%$ and $4.5w/v\%$), this bending is restricted because of enough viscous forces acting on the droplet. At $3w/v\%$, the viscosity is not very high, so there is fast solvent evaporation from the outer shell of the polymer droplet leading to a polymer skin formation outside the droplet. When solvent from the internal core of this droplet gets evaporated, it leads to the collapsing of the polymer droplet, and disc-shaped particles are formed.[14,16] However, at $4.5w/v\%$, the concentration gradient from the skin to the core is comparatively high. In this case, relatively more uniform solvent evaporation occurs from the shell to the core of the jetted droplet leading to the formation of spheres. In the next step, surface-initiated ATRP (SIATRP) is carried out for one hour to grow poly(DMAEMA) brushes from the surface of the spherical, cup, and disc-shaped particles.[14] Brightfield Images and confocal laser scanning microscope (CLSM) images for brush modified and 'as jetted' spheres, cup-shaped particles, and disc-shaped particles are shown in Figs. 12 and 13, respectively.[14,16]

Later α-glucosidase enzyme is allowed to adsorb onto the surface of these brush-modified particles of various shapes via electrostatic interaction.[16] It is found that brush-modified disc-shaped particles exhibit the highest enzyme loading capacity ($58,900\ U/g$ or $589\ mg/g$ of particles), as displayed in Fig. 14. This is in excellent agreement with the simulation results (exhibited in Figs. 9, 10, and 11). The brush-modified spherical particles demonstrate the lowest enzyme immobilization capacity ($41,000\ U/g$ or $410\ mg/g$ of particles) (black curve). On the other hand, the brush-modified cup-shaped particles adsorbed (red curve) in between spheres and discs (44,100 U/g or 441 mg/g of particles) (blue curve in Fig. 14).[16] The cup-shaped particles adsorb slightly more enzyme than spheres (within the error bar), as observed by other researchers also.[18] However, their enzyme adsorption capacity was still less than brush-modified disc-shaped particles. This may be due to the difference in the availability of surface area of these particles for polymer brush growth. Cup-shaped particles have the probability of growing polymer brushes from inside and outside surfaces. However, the grafting density of polymer brushes will be less inside the cup due to less accessibility of monomer (DMAEMA) inside the cup surface. Hence, polymer chain growth will be restricted to the confined concave surface. In contrast, the disc surface should be equally available, leading to enhanced grafting density and maximum enzyme immobilization efficiency.[16] The excellent qualitative agreement of the experimental and simulation results further justifies our simulation model.

## 4. Conclusions

In conclusion, we have utilized the DPD approach to present a generic and robust simulation model for surface fabrication of different-shaped microparticles with surface-initiated ATRP brush grafting. In particular, we have considered the initiator embedded cups, spheres, and flat (rectangular/disc-shaped) surfaces for the modification. We then performed a comparative study of biopolymer adsorption on these brush-modified surfaces to appreciate the critical chemical and physical processes happening at the microscopic level.

Independent of the shape of microparticles, the monomer conversion is enhanced with the addition of initiator concentration on the surfaces for a fixed period, hence the radius of gyration and hydrodynamic radius. The reaction rate kinetics depicted a linear brush growth, as expected for any typical living radical polymerization process, thus validating the chemical kinetics of our simulation model. We found that polymerization reduced the shape factor for all the fabricated surfaces at a fixed initiator concentration. However, we noted a more significant change for the rectangular surface due to its initial structural constraint that advanced to form a more globular shape with polymerization. As illustrated by the radial distribution function, we found the uniform distribution of ATRP brushes near the surface for all the initiator concentrations. Still, they were denser at higher initiator concentrations. The brush density smoothly reduced when moved away from the surface, confirming the brush swelling in the solvent.

Further, we have studied biopolymer adsorption on brush-modified surfaces. We have noticed enhanced biopolymer adsorption with the increase of (i) brush length at a given surface with fixed initiator concentration; (ii) initiator concentration, which triggered the brush-modified surfaces with higher grafting density and hence, a more swelled brush morphology, and (iii) the biopolymer concentration. In addition, we observed biopolymer adsorption for an extended period, allowing the amount of adsorbed biopolymer to reach a steady-state (saturation point) value for most cases studied here. Most importantly, when adsorption reached a saturation point at late times, the flat surface (rectangular/disc-shaped) could adsorb (load) more biopolymers than the other two surfaces having nearly the same adsorption. The experimental results verified the same, considering disk-shaped flat surface particles, cups, and spherical particles.

Finally, we have demonstrated a significantly high biopolymer loading for a longer biopolymer chain length for all the surfaces. Nevertheless, the modified flat surface had the maximum adsorption like in other cases. When the small biopolymer molecules were coarse-grained as a single bead, we marked a significantly low and nearly similar loading for all the modified surfaces. Overall, our simulation data appeared to agree with the experimental results. These results could lead to a different way of regulating the shape of complex soft materials crucial to diverse applications in the biomedical field. For example, the shape and surface chemistry are deciding factors in advanced drug delivery and the development of biomedical devices using cell-material interactions.

**Conflicts of interest**

There are no conflicts of interest to declare.

**Acknowledgments**

S.S. thanks to IIT (BHU) for financial support. A.S. acknowledges Science and Engineering Research Board (SERB) (Grant No. ECR/2017/002529), and S.S. acknowledges the Indian Council of Medical Research (ICMR) (Grant No. ICMR/2020-5642) for funding.

**References:**

1     J. W. Yoo, D. J. Irvine, D. E. Discher and S. Mitragotri, *Nature Reviews Drug Discovery*, 2011, **10**, 521–535.
2     S. Sacanna and D. J. Pine, *Current Opinion in Colloid and Interface Science*, 2011, **16**, 96–102.
3     S. Saha, D. Copic, S. Bhaskar, N. Clay, A. Donini, A. J. Hart and J. Lahann, Chemically Controlled Bending of Compositionally Anisotropic Microcylinders, *Angewandte Chemie International Edition*, 2012, **51**, 660–665.
4     S. Rahmani, S. Saha, H. Durmaz, A. Donini, A. C. Misra, J. Yoon and J. Lahann, Chemically Orthogonal Three-Patch Microparticles, *Angewandte Chemie International Edition*, 2014, **53**, 2332–2338.
5     P. Decuzzi, B. Godin, T. Tanaka, S. Y. Lee, C. Chiappini, X. Liu and M. Ferrari, Size and shape effects in the biodistribution of intravascularly injected particles, *Journal of Controlled Release*, 2010, **141**, 320–327.
6     A. B. Jindal, The effect of particle shape on cellular interaction and drug delivery applications of micro- and nanoparticles, *International Journal of Pharmaceutics*, 2017, **532**, 450–465.
7     B. R. Smith, P. Kempen, D. Bouley, A. Xu, Z. Liu, N. Melosh, H. Dai, R. Sinclair and S. S. Gambhir, Shape matters: Intravital microscopy reveals surprising geometrical dependence for nanoparticles in tumor models of extravasation, *Nano Letters*, 2012, **12**, 3369–3377.
8     V. P. Chauhan, Z. Popović, O. Chen, J. Cui, D. Fukumura, M. G. Bawendi and R. K. Jain, Fluorescent Nanorods and Nanospheres for Real-Time In Vivo Probing of Nanoparticle Shape-Dependent Tumor Penetration, *Angewandte Chemie International Edition*, 2011, **50**, 11417–11420.

9	R. A. Meyer, M. P. Mathew, E. Ben-Akiva, J. C. Sunshine, R. B. Shmueli, Q. Ren, K. J. Yarema and J. J. Green, Anisotropic biodegradable lipid coated particles for spatially dynamic protein presentation, *Acta Biomaterialia*, 2018, **72**, 228–238.

10	W. Li, T. Suzuki and H. Minami, The interface adsorption behavior in a Pickering emulsion stabilized by cylindrical polystyrene particles, *Journal of Colloid and Interface Science*, 2019, **552**, 230–235.

11	B. Heidarshenas, H. Wei, Z. A. Moghimi Moghimi, G. Hussain, F. Baniasadi and G. Naghieh, Nanowires in magnetic drug targeting, *Material Science & Engineering International Journal*, 2019, **3**, 3–9.

12	I. Mirza and S. Saha, Biocompatible Anisotropic Polymeric Particles: Synthesis, Characterization, and Biomedical Applications, *ACS Applied Bio Materials*, 2020, **3**, 8241–8270.

13	M. Alvarez-Paino, M. H. Amer, A. Nasir, V. Cuzzucoli Crucitti, J. Thorpe, L. Burroughs, D. Needham, C. Denning, M. R. Alexander, C. Alexander and F. R. A. J. Rose, Polymer Microparticles with Defined Surface Chemistry and Topography Mediate the Formation of Stem Cell Aggregates and Cardiomyocyte Function, *ACS Applied Materials & Interfaces*, 2019, **11**, 34560–34574.

14	Ifra and S. Saha, Fabrication of topologically anisotropic microparticles and their surface modification with pH responsive polymer brush, *Materials Science and Engineering*, 2019, **104**, 109894.

15	Ifra, A. Singh and S. Saha, Shape Shifting of Cup Shaped Particles on Growing poly (2-hydroxy ethyl methacrylate) Brushes by "Grafting From" Approach and Dissipative Particle Dynamics Simulation, *ChemistrySelect*, 2020, **5**, 4685–4694.

16	Ifra, A. Singh and S. Saha, High Adsorption of α-Glucosidase on Polymer Brush-Modified Anisotropic Particles Acquired by Electrospraying - A Combined Experimental and Simulation Study, *ACS Applied Bio Materials*, 2021, **4**, 7431–7444.

17	S. S. Pradhan and S. Saha, Advances in design and applications of polymer brush modified anisotropic particles, *Advances in Colloid and Interface Science*, 2022, **300**, 102580.

18	J. Chen, V. Kozlovskaya, A. Goins, J. Campos-Gomez, M. Saeed and E. Kharlampieva, Biocompatible Shaped Particles from Dried Multilayer Polymer Capsules, *Biomacromolecules*, 2013, **14**, 3830–3841.

19	H. Kim, H. Terazono, H. Takei and K. Yasuda, Cup-Shaped Superparamagnetic Hemispheres for Size-Selective Cell Filtration, *Scientific Reports*, 2014, **4**, 1–6.

20	Z. Tong and Y. Deng, Synthesis of polystyrene encapsulated nanosaponite composite latex via miniemulsion polymerization, *Polymer*, 2007, **48**, 4337–4343.

21	S. Jairam, Z. Tong, L. Wang and B. Welt, Encapsulation of a Biobased Lignin–Saponite Nanohybrid into Polystyrene Co-Butyl Acrylate (PSBA) Latex via Miniemulsion Polymerization, *ACS Sustainable Chemistry and Engineering*, 2013, **1**, 1630–1637.

22	L. Wang, Y. Liu, J. He, M. J. Hourwitz, Y. Yang, J. T. Fourkas, X. Han, Z. Nie, L. Wang, Y. Liu, J. He, M. J. Hourwitz, Y. Yang, Z. Nie, X. Han and J. T. Fourkas, Continuous Microfluidic Self-Assembly of Hybrid Janus-Like Vesicular Motors: Autonomous Propulsion and Controlled Release, *Small*, 2015, **11**, 3762–3767.

23	S. Li, X. Yang, S. Yang, M. Zhu and X. Wang, TECHNOLOGY PROSPECTING ON ENZYMES: APPLICATION, MARKETING AND ENGINEERING, *Computational and Structural Biotechnology Journal*, 2012, **2**, e201209017.

24  S. Raveendran, B. Parameswaran, S. B. Ummalyma, A. Abraham, A. K. Mathew, A. Madhavan, S. Rebello and A. Pandey, Applications of Microbial Enzymes in Food Industry, *Food Technology and Biotechnology*, 2018, **56**, 16–30.

25  S. Bhatia, in *Introduction to Pharmaceutical Biotechnology, Volume 2*, IOP Publishing, 2018.

26  D. Wang and W. Jiang, Preparation of chitosan-based nanoparticles for enzyme immobilization, *International Journal of Biological Macromolecules*, 2019, **126**, 1125–1132.

27  D. M. Liu, J. Chen and Y. P. Shi, α-Glucosidase immobilization on chitosan-modified cellulose filter paper: Preparation, property and application, *International Journal of Biological Macromolecules*, 2019, **122**, 298–305.

28  Y. K. Cen, Y. X. Liu, Y. P. Xue and Y. G. Zheng, Immobilization of Enzymes in/on Membranes and their Applications, *Advanced Synthesis & Catalysis*, 2019, **361**, 5500–5515.

29  A. T. Thodikayil, S. Sharma and S. Saha, Engineering Carbohydrate-Based Particles for Biomedical Applications: Strategies to Construct and Modify, *ACS Applied Bio Materials*, 2021, **4**, 2907–2940.

30  N. R. Mohamad, N. H. C. Marzuki, N. A. Buang, F. Huyop and R. A. Wahab, An overview of technologies for immobilization of enzymes and surface analysis techniques for immobilized enzymes, *Agriculture and Environmental Biotechnology*, 2015, **29**, 205–220.

31  G. F. D. del Castillo, M. Koenig, M. Müller, K. J. Eichhorn, M. Stamm, P. Uhlmann and A. Dahlin, Enzyme Immobilization in Polyelectrolyte Brushes: High Loading and Enhanced Activity Compared to Monolayers, *Langmuir*, 2019, **35**, 3479–3489.

32  B. Haupt, T. Neumann, A. Wittemann and M. Ballauff, Activity of Enzymes Immobilized in Colloidal Spherical Polyelectrolyte Brushes, *Biomacromolecules*, 2005, **6**, 948–955.

33  J. Zdarta, A. S. Meyer, T. Jesionowski and M. Pinelo, A General Overview of Support Materials for Enzyme Immobilization: Characteristics, Properties, Practical Utility, *Catalysts*, 2018, **8**, 92.

34  B. Krajewska, Application of chitin- and chitosan-based materials for enzyme immobilizations: a review, *Enzyme and Microbial Technology*, 2004, **35**, 126–139.

35  R. Li, / Obc, M. Hoarau, S. Badieyan, E. Neil and G. Marsh, Immobilized enzymes: understanding enzyme – surface interactions at the molecular level, *Organic & Biomolecular Chemistry*, 2017, **15**, 9539–9551.

36  C. Marschelke, I. Raguzin, A. Matura, A. Fery and A. Synytska, Controlled and tunable design of polymer interface for immobilization of enzymes: does curvature matter?, *Soft Matter*, 2017, **13**, 1074–1084.

37  C. Marschelke, M. Müller, D. Köpke, A. Matura, M. Sallat and A. Synytska, Hairy Particles with Immobilized Enzymes: Impact of Particle Topology on the Catalytic Activity, *ACS Applied Materials and Interfaces*, 2019, **11**, 1645–1654.

38  A. Kusumo, L. Bombalski, Q. Lin, K. Matyjaszewski, J. W. Schneider and R. D. Tilton, High Capacity, Charge-Selective Protein Uptake by Polyelectrolyte Brushes, *Langmuir*, 2007, **23**, 4448–4454.

39  P. Jain, J. Dai, S. Grajales, S. Saha, G. L. Baker and M. L. Bruening, Completely aqueous procedure for the growth of polymer brushes on polymeric substrates, *Langmuir*, 2007, **23**, 11360–11365.

40  R. D. Groot and P. B. Warren, Dissipative particle dynamics: Bridging the gap between atomistic and mesoscopic simulation, *Journal of Chemical Physics*, 1997, **107**, 4423–4435.

41  P. Espanol and P. Warren, Statistical-Mechanics of Dissipative Particle Dynamics, *Europhysics Letters*, 1995, **30**, 191–196.

42  A. Singh, O. Kuksenok, J. A. Johnson and A. C. Balazs, Tailoring the structure of polymer networks with iniferter-mediated photo-growth, *Polym. Chem.*, 2016, **7**, 2955–2964.

43  X. Yong, O. Kuksenok, K. Matyjaszewski and A. C. Balazs, Harnessing interfacially-active nanorods to regenerate severed polymer gels, *Nano Letters*, 2013, **13**, 6269–6274.

44  A. G. Schlijper, P. J. Hoogerbrugge and C. W. Manke, Computer simulation of dilute polymer solutions with the dissipative particle dynamics method, *Journal of Rheology*, 1998, **39**, 567.

45  N. A. Spenley, Scaling laws for polymers in dissipative particle dynamics, *Europhysics Letters*, 2000, **49**, 534.

46  M. B. Liu, G. R. Liu, L. W. Zhou and J. Z. Chang, Dissipative Particle Dynamics (DPD): An Overview and Recent Developments, *Archives of Computational Methods in Engineering*, 2014, **22**, 529–556.

47  Y. H. Feng, Y. H. Feng, X. P. Zhang, X. P. Zhang, Z. Q. Zhao, Z. Q. Zhao, X. D. Guo and X. D. Guo, Dissipative Particle Dynamics Aided Design of Drug Delivery Systems: A Review, *Molecular Pharmaceutics*, 2020, **17**, 1778–1799.

48  J. Wang, Y. Han, Z. Xu, X. Yang, S. Ramakrishna and Y. Liu, Dissipative Particle Dynamics Simulation: A Review on Investigating Mesoscale Properties of Polymer Systems, *Macromolecular Materials and Engineering*, 2021, **306**, 2000724.

49  P. Espanol and P. B. Warren, *Journal of Chemical Physics*, 2017, 146, 150901.

50  M. B. Liu, G. R. Liu, L. W. Zhou and J. Z. Chang, Dissipative Particle Dynamics (DPD): An Overview and Recent Developments, *Archives of Computational Methods in Engineering*, 2015, **22**, 529–556.

51  P. Nikunen, M. Karttunen and I. Vattulainen, How would you integrate the equations of motion in dissipative particle dynamics simulations?, *Computer Physics Communications*, 2003, **153**, 407–423.

52  R. D. Groot, A Local Galilean Invariant Thermostat, *Journal of Chemical Theory and Computation*, 2006, **2**, 568–574.

53  P. J. Hoogerbrugge and J. M. V. A. Koelman, Simulating Microscopic Hydrodynamic Phenomena with Dissipative Particle Dynamics, *Europhysics Letters*, 1992, **19**, 155–160.

54  S. Plimpton, Fast Parallel Algorithms for Short-Range Molecular-Dynamics, *Journal of Computational Physics*, 1995, **117**, 1–19.

55  V. Symeonidis, G. E. Karniadakis and B. Caswell, Schmidt number effects in dissipative particle dynamics simulation of polymers, *The Journal of Chemical Physics*, 2006, **125**, 184902.

56  C. Junghans, M. Praprotnik and K. Kremer, Transport properties controlled by a thermostat: An extended dissipative particle dynamics thermostat, *Soft Matter*, 2008, **4**, 156–161.

57    A. K. Singh, A. Chauhan, S. Puri and A. Singh, Photo-induced bond breaking during phase separation kinetics of block copolymer melts: a dissipative particle dynamics study, *Soft Matter*, 2021, **17**, 1802–1813.
58    H.-P. Hsu and K. Kremer, Static and dynamic properties of large polymer melts in equilibrium, *The Journal of Chemical Physics*, 2016, **144**, 154907.
59    Y. Liu, O. Kuksenok, X. He, M. Aizenberg, J. Aizenberg and A. C. Balazs, Harnessing Cooperative Interactions between Thermoresponsive Aptamers and Gels to Trap and Release Nanoparticles, *ACS Applied Materials and Interfaces*, 2016, **8**, 30475–30483.
60    X. Yong, O. Kuksenok and A. C. Balazs, Modeling free radical polymerization using dissipative particle dynamics, *Polymer*, 2015, **72**, 217–225.
61    J. Genzer, In silico polymerization: Computer simulation of controlled radical polymerization in bulk and on flat surfaces, *Macromolecules*, 2006, **39**, 7157–7169.
62    H. F. Gao, K. Min and K. Matyjaszewski, Gelation in ATRP Using Structurally Different Branching Reagents: Comparison of Inimer, Divinyl and Trivinyl Cross-Linkers, *Macromolecules*, 2009, **42**, 8039–8043.
63    A. Singh, O. Kuksenok, J. A. Johnson and A. C. Balazs, Photo-regeneration of severed gel with iniferter-mediated photo-growth, *Soft Matter*, 2017, **13**, 1978–1987.
64    S. Biswas, A. Singh, A. Beziau, T. Kowalewski, K. Matyjaszewski and A. C. Balazs, Modeling the formation of layered, amphiphilic gels, *Polymer*, 2017, **111**, 214–221.
65    K. Matyjaszewski and J. H. Xia, Atom transfer radical polymerization, *Chemical Reviews*, 2001, **101**, 2921–2990.
66    J. S. Wang and K. Matyjaszewski, Controlled Living Radical Polymerization - Atom-Transfer Radical Polymerization in the Presence of Transition-Metal Complexes, *J Am Chem Soc*, 1995, **117**, 5614–5615.
67    A. Beziau, A. Singh, R. N. L. de Menezes, H. Ding, A. Simakova, O. Kuksenok, A. Balazs, T. Kowalewski and K. Matyjaszewski, Miktoarm star copolymers as interfacial connectors for stackable amphiphilic gels, *Polymer*, 2016, **101**, 406–414.
68    F. J. M. de Meyer, M. Venturoli and B. Smit, Molecular Simulations of Lipid-Mediated Protein-Protein Interactions, *Biophysical Journal*, 2008, **95**, 1851–1865.
69    R. D. Groot and K. L. Rabone, Mesoscopic Simulation of Cell Membrane Damage, Morphology Change and Rupture by Nonionic Surfactants, *Biophysical Journal*, 2001, **81**, 725–736.
70    Y. L. Wang, Z. Y. Lu and A. Laaksonen, Specific binding structures of dendrimers on lipid bilayer membranes, *Physical Chemistry Chemical Physics*, 2012, **14**, 8348–8359.
71    A. Vishnyakov, D. S. Talaga and A. v. Neimark, DPD simulation of protein conformations: From α-helices to β-structures, *Journal of Physical Chemistry Letters*, 2012, **3**, 3081–3087.
72    A. C. C. Esteves, K. Lyakhova, L. G. J. van der Ven, R. A. T. M. van Benthem and G. de With, Surface Segregation of Low Surface Energy Polymeric Dangling Chains in a Cross-Linked Polymer Network Investigated by a Combined Experimental-Simulation Approach, *Macromolecules*, 2013, **46**, 1993–2002.
73    Y. Liu, G. T. McFarlin, X. Yong, O. Kuksenok and A. C. Balazs, Designing Composite Coatings That Provide a Dual Defense against Fouling, *Langmuir*, 2015, **31**, 7524–7532.
74    B. M. Tande, N. J. Wagner, M. E. Mackay, C. J. Hawker and M. Jeong, Viscosimetric, Hydrodynamic, and Conformational Properties of Dendrimers and Dendrons, *Macromolecules*, 2001, **34**, 8580–8585.

75  M. Baalousha, F. V. D. Kammer, M. Motelica-Heino, H. S. Hilal and P. le Coustumer, Size fractionation and characterization of natural colloids by flow-field flow fractionation coupled to multi-angle laser light scattering, *Journal of Chromatography A*, 2006, **1104**, 272–281.

76  M. Schmidt and W. Burchard, Translational diffusion and hydrodynamic radius of unperturbed flexible chains, *Macromolecules*, 2002, **14**, 210–211.

77  W. Burchard, Solution Properties of Branched Macromolecules, *Advances in Polymer Science*, 1999, **143**, 113–194.

**Figures and Table**

Table

| $a_{ij}$ | w | s | m | M | B | b |
|---|---|---|---|---|---|---|
| Wall ($w$) | 25 | 60 | 60 | 60 | 60 | 60 |
| Solvent ($s$) | | 25 | 45 | 25 | 25 | 25 |
| Microparticle ($m$) | | | 25 | 35 | 35 | 20 |
| Monomer ($M$) | | | | 25 | 25 | 25 |
| Brush ($B$) | | | | | 27 | 15 |
| Biopolymer ($b$) | | | | | | 27 |

Table 1: The interaction parameter $a_{ij}$ used in the simulation for different DPD beads.

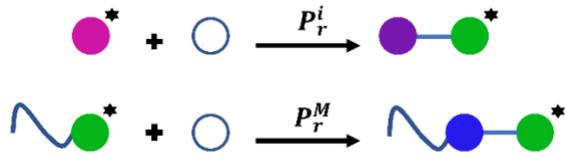

**Figure 1**: Schematic diagram illustrating the atom transfer radical polymerization (ATRP) process from the initiators embedded at the nanoparticle surface.

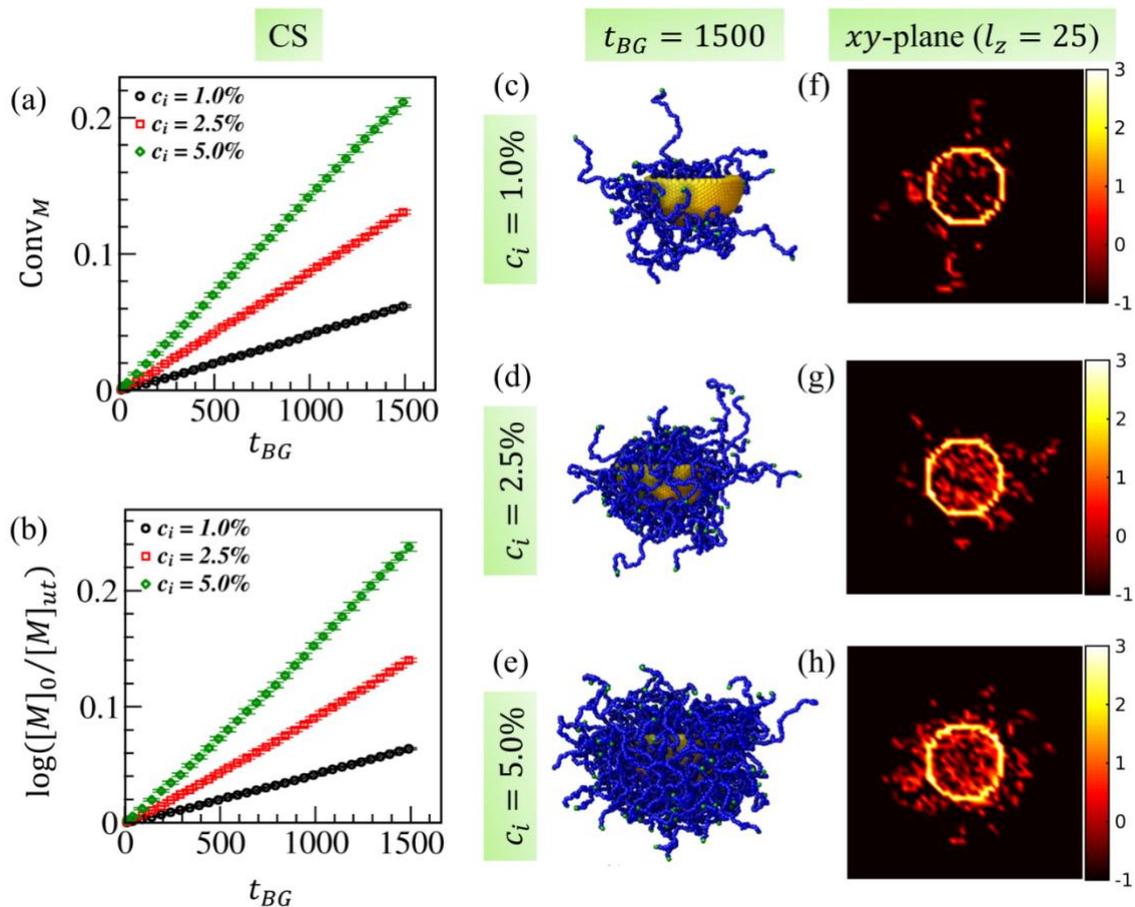

**Figure 2**: (a) Monomer conversion ($\text{Conv}_M$) as a function of time ($t_{BG}$) for different initiator concentrations: $c_i = 1.0\%$ (black curve), 2.5% (red curve), and 5.0% (green curve) presented at CS. (b) Temporal change in the monomer conversion rate, $\log([M]_0/[M]_{ut})$ for the reaction kinetics corresponding to data in (a). (c-e) The ATRP brush modified cup surfaces at different $c_i$ for the monomer conversion up to $t_{BG} = 1500$. (f-h) Crosssection ($xy$-plane) plots at $l_z = 25$ corresponding to the brush modified CS in (c), (d), and (e), respectively, indicating the brush density variation around CS. The color bar is at the extreme right.

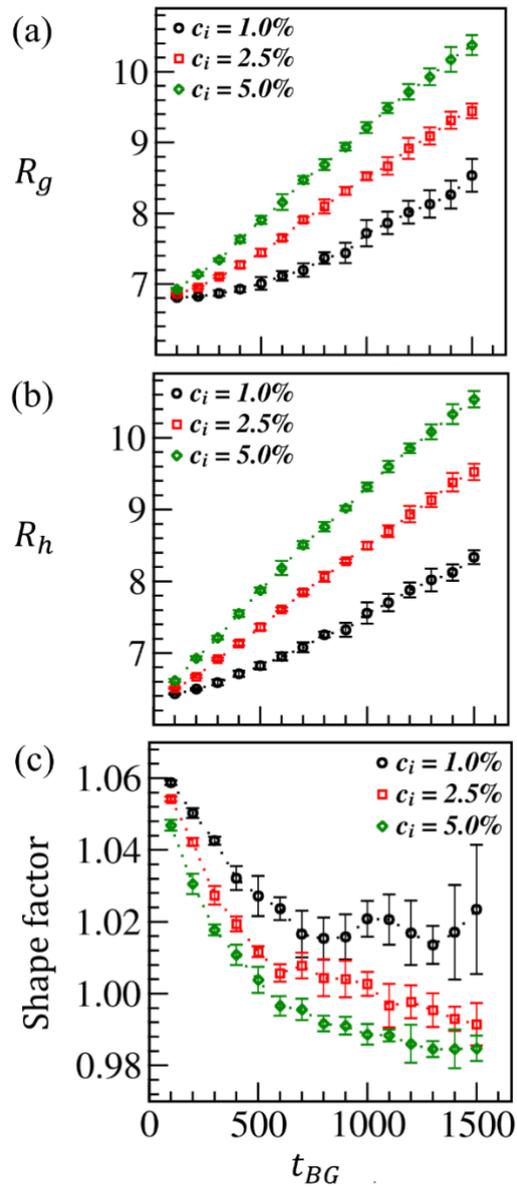

**Figure 3**: Variation in the radius of gyration ($R_g$) in (a), the hydrodynamic radius ($R_h$) in (b), and the shape factor ($\rho_{sf} = R_g/R_h$) against ATRP brush growth time $t_{BG}$ for CS. The black, red, and green curves illustrate $R_g$ for the initiator concentration, $c_i = 1.0\%$, 2.5%, and 5.0%, respectively, embedded at CS.

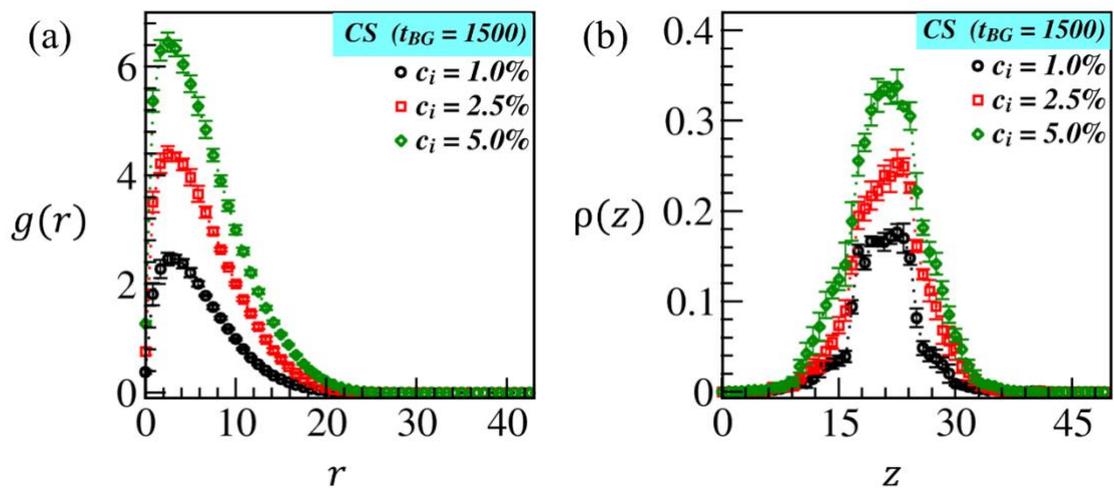

**Figure 4**: (a) Radial distribution function (RDF): $g(r)$ against radial distance $r$, and (b) density distribution: $\rho(z)$ along $z$-direction of the grown ATRP-brushes at $t_{BG} = 1500$ at CS for different initiator concentrations $c_i = 1.0\%$ (in black symbol), 2.5% (in red symbol) and 5.0% (in green symbol) corresponding to the snapshots shown in Figs. 2(c-e).

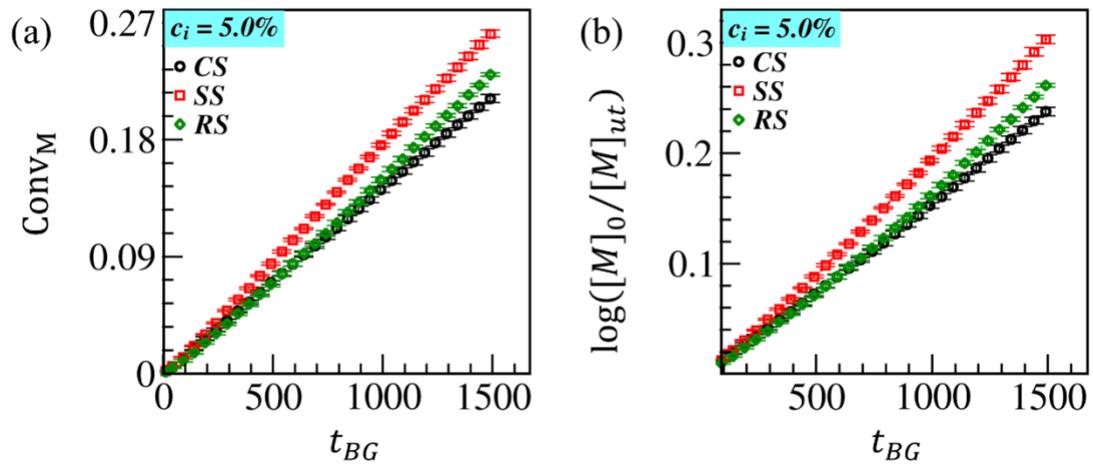

**Figure 5**: Comparison of temporal variation of $\text{Conv}_M$ in (a), and $\log([M]_0/[M]_{ut})$ in (b) at the cup surface (CS) in black, spherical surface (SS) in red, and rectangular surface (RS) in green curves for $c_i = 5.0\%$ embedded on their surfaces.

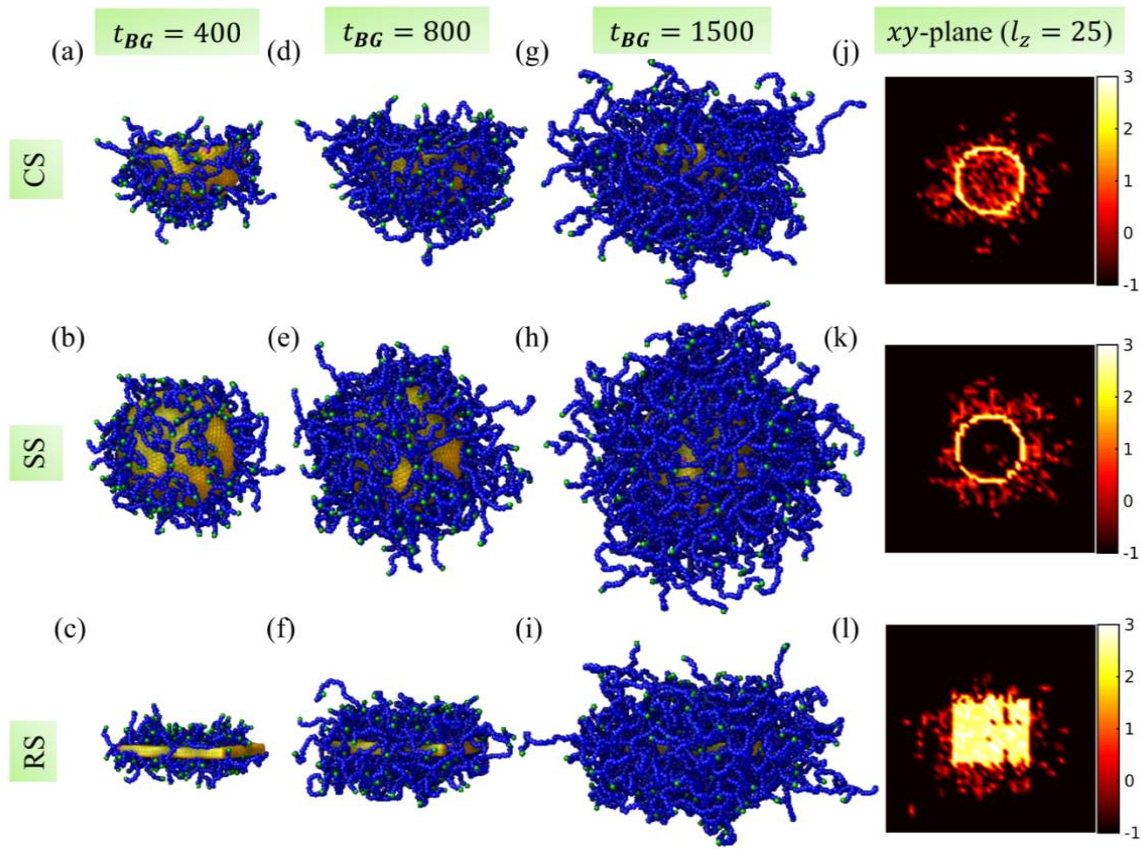

**Figure 6**: ATRP brush modified CS, SS, and RS for $c_i = 5.0\%$ and for the brush growth (polymerization) time up to $t_{BG} = 400$ in (a-c), 800 in (d-f), and 1500 in (g-i). (j-l) demonstrate the brush density variation in the $xy$ cross-section at $l_z = 25$ corresponding to the snapshots in (g-i).

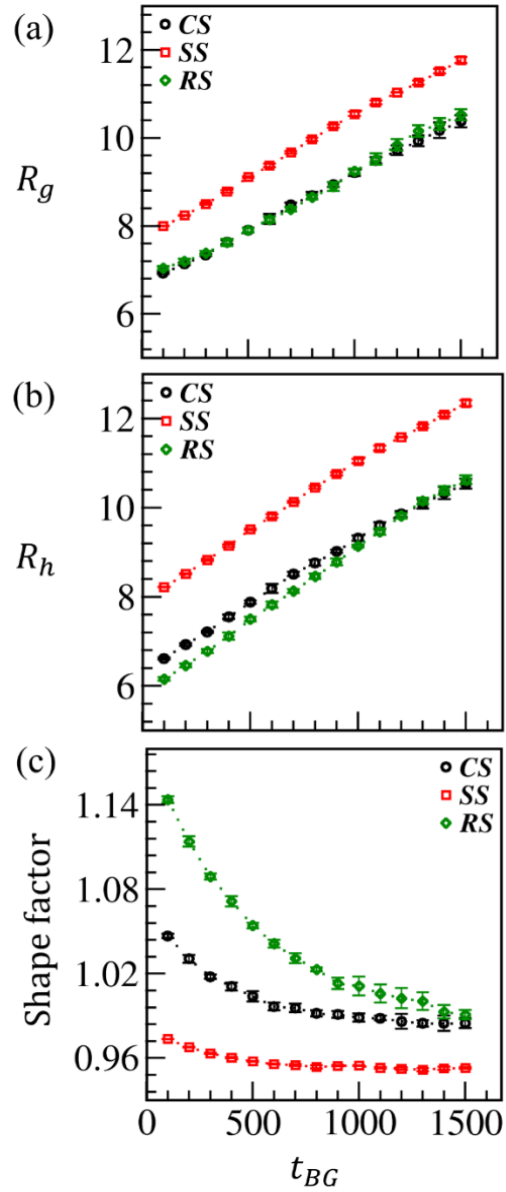

**Figure 7**: Comparison of (a) radius of gyration ($R_g$), (b) hydrodynamic radius ($R_h$), and (c) size factor ($\rho_{sf}$) against the polymerization time $t_{BG}$ for the brush modified CS (black curve), SS (red curve), and RS (green curve), respectively, at $c_i = 5\%$.

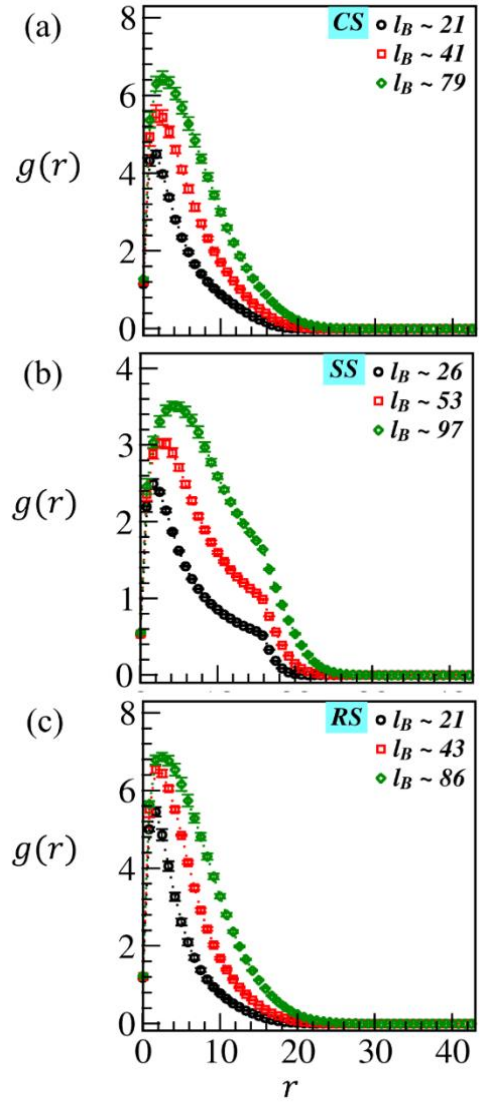

**Figure 8**: Comparison of RDF ($g(r)$) of ATRP brushes around (a) CS, (b) SS, and (c) RS at different brush lengths observed at different polymerization times for $c_i = 5\%$.

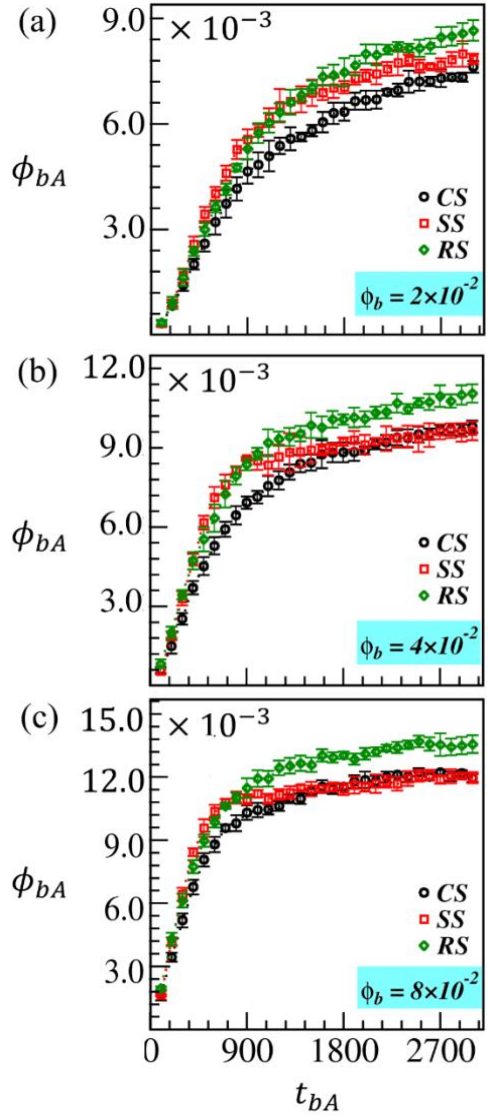

**Figure 9**: The estimated fraction of adsorbed biopolymers ($\phi_{bA}$) on the brush modified CS (black symbols), SS (red symbols), and RS (green symbols) in the presence of a different fraction of biopolymers (a) $\phi_b = 2 \times 10^{-2}$, (b) $\phi_b = 4 \times 10^{-2}$ and (c) $\phi_b = 8 \times 10^{-2}$, respectively, in the solution for $c_i = 5\%$. The adsorption process is monitored for at $t_{bA} = 3000$.

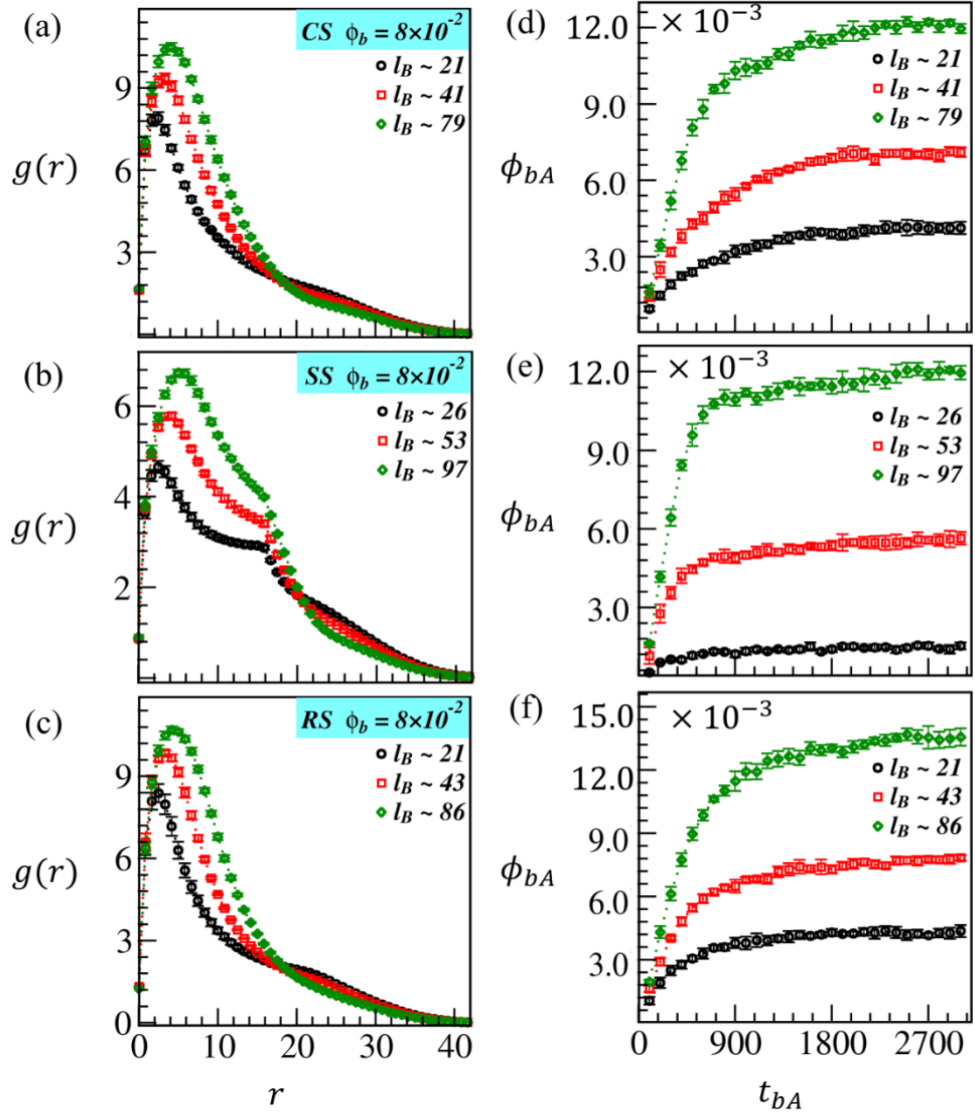

**Figure 10**: Comparing the biopolymer adsorption for different brush lengths grew up to $t_{BG} = 400$, 800, and 1500 on CS, SS, and RS, respectively. Corresponding brush lengths are displayed in the legend. For each brush length, the adsorption is considered till $t_{bA} = 3000$. (a-c) Show $g(r)$ versus $r$ for the biopolymers around CS, SS, and RS beads, respectively. (d-f) Display the temporal variation in the adsorbed fraction of biopolymers ($\phi_{bA}$).

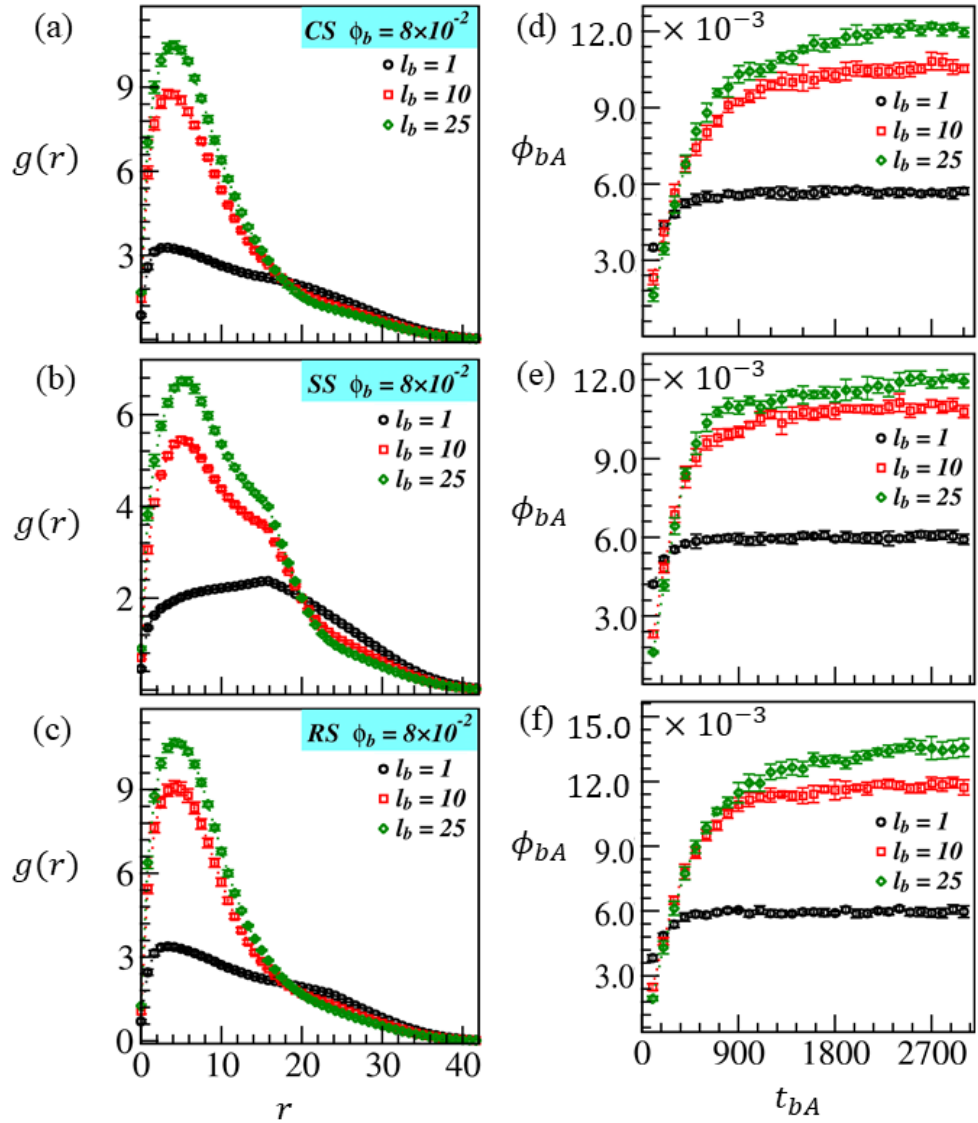

**Figure 11**: Assessment of biopolymer adsorption of different chain lengths: $l_b = 1$ (black curve), $l_b = 10$ (red curve), and $l_b = 25$ (green curve) on the brush modified surfaces: CS, SS, and RS. The adsorption is allowed to continue for $t_{bA} = 3000$. (a-c) RDF plot ($g(r)$ versus $r$), and (d-f) the temporal variation of the adsorbed fraction of biopolymers ($\phi_{bA}$).

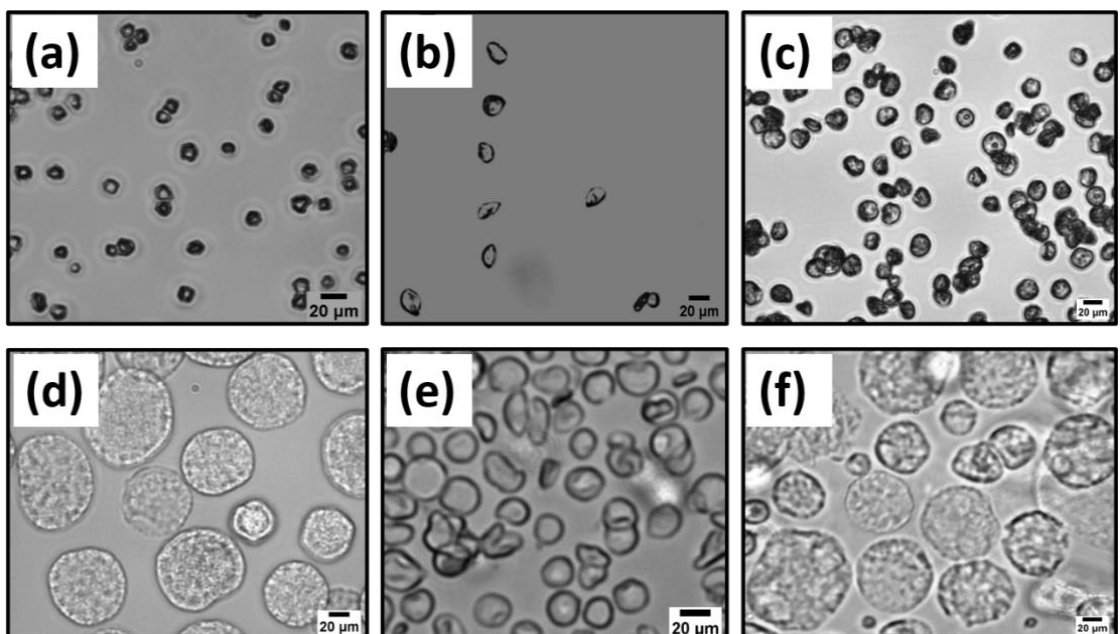

**Figure 12**: Brightfield images of poly(DMAEMA) brush unmodified (a) spheres (average size – 4.9 μm ± 1.1) (b) cups (average size – 5.9 μm ± 1.6) (c) discs (average size – 8.4 μm ± 1.0) and poly(DMAEMA) brush modified (d) spheres (average size – 16.2 μm ± 3.2) (e) cup-shaped particles (average size - 9.2 μm ± 1.8) (f) disc-shaped particles (average size – 18.4 μm ± 2.9).

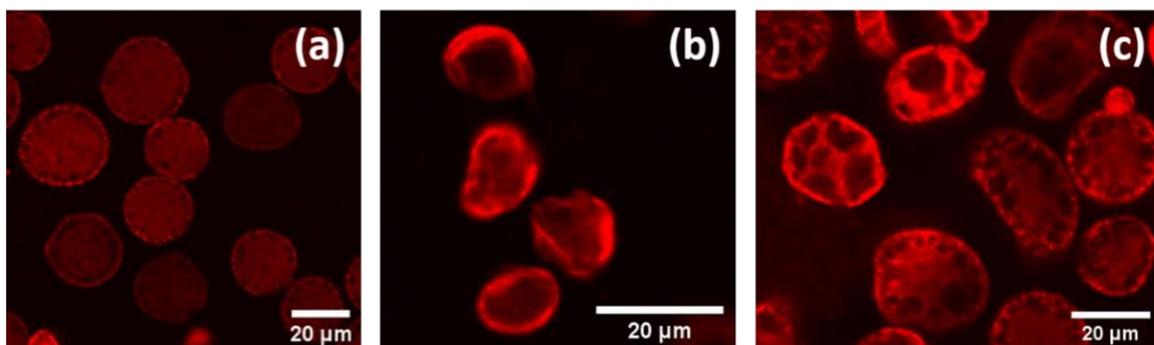

**Figure 13**: Confocal laser scanning microscopic (CLSM) images of poly(DMAEMA) brush modified (a) spheres (average size – 17.1 μm ± 3.6) (b) cup-shaped particles (average size – 8.9 μm ± 1.4) (c) disc-shaped particles (average size – 19.2 μm ± 3.9).

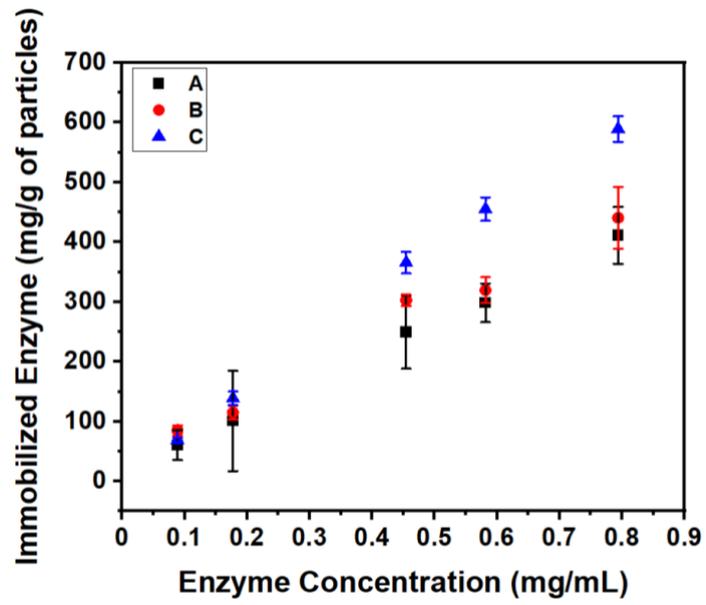

**Figure 14**: Immobilized enzyme (mg) per g of particles (A) spheres (B) cup-shaped particles (C) disc-shaped particles.

# Supporting information
# Dissipative particle dynamics simulation study on ATRP-brush modification of variably shaped surfaces and biopolymer adsorption

Samiksha Shrivastava[1], Ifra[2], Sampa Saha[2], and Awaneesh Singh[1*]

[1]Department of Physics, Indian Institute of Technology (BHU), Varanasi, India

[2]Department of Materials Science and Engineering, Indian Institute of Technology Delhi, New Delhi, India

1. Experimental procedure

    1.1. Materials

   Polylactide (3052D, $M_n$ - 116,000 g/mol) is obtained from Nature Work. Methyl Methacrylate (>99.0%), 2-hydroxy ethyl methacrylate (>95.0%), 2-dimethyl aminoethyl methacrylate (>98.5%) are procured from TCI. The acrylate monomers are passed through a column of basic alumina to remove the inhibitor. Copper (I) bromide is supplied from Spectrochem, and bromopropionyl bromide are procured by TCI. Copper bromide (I) is washed with glacial acetic acid and ethanol, respectively, for purification. Fischer Scientific supplies basic alumina neutral alumina, dimethylformamide, and ethylene diamine tetraacetate. Chloroform and Tween-20 are procured from Merck. Methyl bromo propionate and Pentadiethylene tetraacetate (PMDETA) are obtained from Sigma Aldrich. α-Glucosidase (Maltase) extracted from yeast (which contained 100U (active)/mg), p-nitrophenyl-α-D-glucopyranoside (α-PNPG, extra pure, 98%), and Sisco Research Laboratories supply glutathione. Rhodamine B dye is purchased from India Mart.

    1.2. Fabrication of spherical, cup-shaped, and disc-shaped particles and their surface modification by growing Poly(DMAEMA) Brushes

   First, copolymerization of methyl methacrylate and 2-hydroxy ethyl methacrylate (HEMA) is carried out by the bulk ATRP method. The hydroxyl groups (in HEMA unit) are converted to ATRP-initiating moiety (BEMA unit) by reacting them with bromopropionyl bromide to yield poly($MMA_{0.9}$-co-$BEMA_{0.1}$) ($M_n$: 9155 g/mol; PDI:1.38) as discussed in our previous publication.[1] Then, spherical, cup-shaped, and disc-shaped particles are made via the electrohydrodynamic jetting (EHDJ) technique. A polymer solution is obtained by a blend of 75 wt % PLA and 25 wt % poly(MMA-co-BEMA) dissolved in a solvent system (chloroform: DMF - 97:3). This is sprayed at different concentrations, flow rates, and applied voltage (details of solution/processing parameters are given in Table S1). Then, poly(DMAEMA) chains (polymer brushes) are grown from the surface of spherical, cup-shaped, and disc-shaped particles using the "grafting from" surface-initiated atom transfer radical polymerization

(SIATRP) as discussed in our previous paper.[1] Except spherical, cup and disc-shaped particle formation and brush modifications have already been discussed in our earlier publications.[1,2]

| Shape | Concentration of polymer solution (w/v%) ((75%PLA +25%poly(MMA-co-BEMA)) | Flow rate (ml/h) | Voltage (kV) |
|---|---|---|---|
| Spheres | 4.5 | 0.5 | 8.5 |
| Cups | 1 | 1 | 9 |
| Discs | 3 | 1.5 | 9 |

Table S1: Solution/Processing parameters to fabricate spheres, cups, and discs.

### 1.3. Immobilization of the α-Glucosidase Enzyme onto the brush modified spherical, cup-shaped, and disc-shaped particles[2]

First, α-glucosidase is mixed in a solvent (phosphate buffer solution (pH ~ 6.8)) at a concentration of 0.8 mg/mL or 80 U/mL (one unit (U) enzyme may be defined as the amount which can catalyze the conversion of 1 μmol of α-PNPG in a minute)[2]. Initially, particles (1mg) are dispersed in 100 μL (0.08mg/mL) in a 1.5 mL centrifuge tube, and particles are stirred on a plate shaker for 1 h at room temperature. After 1h, particles are collected in a centrifuge at 10,000 rpm and washed three times. The immobilized enzyme amount (active units) is obtained by deducting the amount of the enzyme present in the supernatants (initial supernatants and supernatant obtained on washing the particles) from the initial solution. Then, α-PNPG (substrate), which gets hydrolyzed by the α-glucosidase enzyme to release p-nitrophenol (p-NP), is taken for enzyme activity determination.[2] Thus, the amount (mmoles) of released *p*-NP is identified by UV/vis spectroscopy at 400 nm (a calibration curve is already prepared using pure *p*-nitrophenol)[2]. The experiments are performed in triplicate.

### 1.4. Characterization Methods.

UV absorbance is done on Synergy h1, microplate reader. CLSM images are taken on a confocal laser scanning microscope (CLSM), Leica LAS 2.6.0 build7266. Bright-field images are taken on an optical microscope (Model -Olympus IX73). The samples for both pictures are prepared by mounting dispersed particles on a glass slide and then covering the drop with coverslips.

### 2. References:

1  Ifra and S. Saha, Fabrication of topologically anisotropic microparticles and their surface modification with pH responsive polymer brush, *Materials Science and Engineering*, 2019, **104**, 109894.
2  Ifra, A. Singh and S. Saha, High Adsorption of α-Glucosidase on Polymer Brush-Modified Anisotropic Particles Acquired by Electrospraying - A Combined Experimental and Simulation Study, *ACS Applied Bio Materials*, 2021, **4**, 7431–7444.